\documentclass[twocolumn]{aastex63}

\usepackage{amsmath}	
\usepackage{amssymb}	
\DeclareRobustCommand{\ion}[2]{%
\relax\ifmmode
\ifx\testbx\f@series
{\mathbf{#1\,\mathsc{#2}}}\else
{\mathrm{#1\,\mathsc{#2}}}\fi
\else\textup{#1\,{\mdseries\textsc{#2}}}%
\fi}

\usepackage{rotating}

\usepackage{graphicx}	
\usepackage{subfigure}
\usepackage{threeparttable}

\received{February 24, 2021}
\revised{May 18, 2021}
\accepted{May 25, 2021}
\submitjournal{AAS Publishing}

\shorttitle{VLASS Source Statistics}
\shortauthors{Gordon et al.}
\graphicspath{{./}{figures/}}

\begin{document}

\title{A \textit{Quick Look} at the $3\,$GHz Radio Sky I. Source Statistics from the Very Large Array Sky Survey}

\correspondingauthor{Yjan Gordon}
\email{yjan.gordon@umanitoba.ca}

\author[0000-0003-1432-253X]{Yjan~A. Gordon}
\affil{Department of Physics and Astronomy, University of Manitoba, 
Winnipeg, MB R3T 2N2, Canada}

\author[0000-0001-5470-3084]{Michelle~M. Boyce}
\affiliation{Department of Physics and Astronomy, University of Manitoba, 
Winnipeg, MB R3T 2N2, Canada}

\author[0000-0001-6421-054X]{Christopher~P. O'Dea}
\affiliation{Department of Physics and Astronomy, University of Manitoba, 
Winnipeg, MB R3T 2N2, Canada}

\author[0000-0001-5636-7213]{Lawrence Rudnick}
\affiliation{Minnesota Institute for Astrophysics, School of Physics and Astronomy, University of Minnesota, 116 Church Street SE,\\ Minneapolis, MN 55455, USA}

\author[0000-0003-4873-1681]{Heinz Andernach}
\affiliation{Departamento de Astronom\'ia, DCNE, Universidad de Guanajuato, Cj\'on de Jalisco s/n, 36023 Guanajuato, GTO, Mexico}

\author[0000-0003-4227-4838]{Adrian~N. Vantyghem}
\affiliation{Department of Physics and Astronomy, University of Manitoba, 
Winnipeg, MB R3T 2N2, Canada}

\author{Stefi~A. Baum}
\affiliation{Department of Physics and Astronomy, University of Manitoba, 
Winnipeg, MB R3T 2N2, Canada}

\author{Jean-Paul Bui}
\affiliation{Dunlap Institute for Astronomy and Astrophysics, University of Toronto, 50 St George Street, Toronto, ON M5S 3H4, Canada}

\author{Mathew Dionyssiou}
\affiliation{Dunlap Institute for Astronomy and Astrophysics, University of Toronto, 50 St George Street, Toronto, ON M5S 3H4, Canada}

\author[0000-0001-6189-7665]{Samar Safi-Harb}
\affil{Department of Physics and Astronomy, University of Manitoba, 
Winnipeg, MB R3T 2N2, Canada}

\author{Isabel Sander}
\affil{Department of Physics and Astronomy, University of Manitoba, 
Winnipeg, MB R3T 2N2, Canada}



\begin{abstract}

The Very Large Array Sky Survey (VLASS) is observing the entire sky north of $-40^{\circ}$ in the S-band ($2<\nu<4\,$GHz), with the highest angular resolution ($2''.5$) of any all-sky radio continuum survey to date.
VLASS will cover its entire footprint over three distinct epochs, the first of which has now been observed in full.
Based on \textit{Quick Look} images from this first epoch, we have created a catalog of $1.9\times10^{6}$ reliably detected radio components.
Due to the limitations of the \textit{Quick Look} images, component flux densities are underestimated by $\sim 15\,\%$ at $S_{\text{peak}}>3\,$mJy/beam and are often unreliable for fainter components.
We use this catalog to perform statistical analyses of the $\nu \sim 3\,$GHz radio sky.
Comparisons with the Faint Images of the Radio Sky at Twenty cm survey (FIRST) show the typical $1.4-3\,$GHz spectral index, $\alpha$, to be $\sim-0.71$.
The radio color-color distribution of point and extended components is explored by matching with FIRST and the LOFAR Two Meter Sky Survey.
We present the VLASS source counts, $dN/dS$, which are found to be consistent with previous observations at $1.4$ and $3\,$GHz.
Resolution improvements over FIRST result in excess power in the VLASS two-point correlation function at angular scales $\lesssim 7''$, and in $18\,\%$ of active galactic nuclei associated with a single FIRST component being split into multi-component sources by VLASS.

\end{abstract}

\keywords{
\href{http://astrothesaurus.org/uat/1338}{Radio Astronomy (1338)},
\href{http://astrothesaurus.org/uat/1343}{Radio Galaxies (1343)},
\href{http://astrothesaurus.org/uat/1356}{Radio Source Catalogs (1356)},
\href{http://astrothesaurus.org/uat/1464}{Sky Surveys (1464)},
}


\section{Introduction}

The past two decades have seen the advent of wide-field continuum imaging surveys of the radio sky.
The advantages of blind surveys over targeted observations include large numbers of objects for which observations are obtained, and the potential to discover new phenomena \citep{Padovani2016, Norris2017}.
The National Radio Astronomy Observatory (NRAO) Very Large Array Sky Survey \citep[NVSS,][]{Condon1998} set the bar for wide-field radio continuum surveys by surveying $80\,\%$ of the sky at $1.4\,$GHz.
With an angular resolution of $\sim 45''$ and a typical rms noise of $450\,\mu$Jy/beam, NVSS has catalogued $\sim 2\times 10^{6}$ radio detections.
The Faint Images of the Radio Sky at Twenty cm survey \citep[FIRST,][]{Becker1995}, also at $1.4\,$GHz, probes deeper than NVSS with an rms of $130\,\mu$Jy/beam, and improved angular resolution of $5''.4$ while covering $25\,\%$ of the sky.
Meanwhile, at low frequency the Tata Institute of Fundamental Research Giant Metre Radio Telescope Sky Survey \citep[TGSS,][]{Intema2017} has mapped $90\,\%$ of the sky at $150\,$MHz with an rms of $5\,$mJy/beam and an angular resolution of $\sim 25''$.


Technological improvements over the past two decades have allowed upgrades to existing radio telescopes such as the Karl G. Jansky Very Large Array (VLA), and the construction of new facilities such as the LOw Frequency ARray \citep[LOFAR,][]{vanHaarlem2013}, the Australian Square Kilometre Array Pathfinder \citep[ASKAP,][]{Johnston2007}, and the MeerKAT telescope \citep{Jonas2009}.
These state-of-the-art facilities allow for sky surveys that probe deeper, with shorter observing times, and with higher resolution than were previously possible, allowing us to build upon the scientific output of the last generation of continuum sky surveys.
For instance, ASKAP is being used to conduct the Rapid ASKAP Continuum Survey \citep[RACS,][]{McConnell2020} and Evolutionary Map of the Universe survey \citep[EMU,][]{Norris2011}, providing $\nu \sim 1\,$GHz coverage of the Southern Sky, the latter down to sensitivities of $\sim 10\,\mu$Jy/beam.
Meanwhile, the LOFAR Two-metre Sky Survey \citep[LoTSS,][]{Shimwell2017} will survey the entire Northern sky at $\sim150\,\text{MHz}$ with a synthesised beam size of $6''$ and an rms of $\sim 70\,\mu$Jy/beam, providing deep, low-frequency observations to complement existing high-frequency data.

Of the current generation of radio continuum imaging surveys, the highest angular resolution will be provided by the VLA Sky Survey \citep[VLASS,][]{Lacy2020}.
With an angular resolution of $\sim2''.5$, VLASS is currently surveying the entire sky North of $-40^{o}$ in the S-band ($2< \nu < 4\,\text{GHz}$, hereafter referred to as $\nu \sim 3\,$GHz).
Such a small beam size has the advantage of revealing morphological structure on smaller angular scales than other surveys \textemdash\ e.g. many sources that appear compact in FIRST or LoTSS may be resolved by VLASS.
Furthermore the resolution of VLASS allows improved study of the true sizes of radio sources in the sky \citep[e.g.,][]{Allen1962, Cotton2018} as well as more reliable identifications of their optical or infrared hosts.
Survey operations for VLASS began in 2017, and observing for the first of three planned epochs was completed in 2019.

In this paper we describe a catalog of radio components produced from the VLASS epoch 1 \textit{Quick Look} imaging.
Here, and throughout, we use the term radio \textit{component} to refer to a single detection by the source finding algorithm rather than radio \textit{source} as, especially at high resolution, a single physical radio \textit{source} may be described by multiple components.
For example, the two hot spots of a double-lobed radio galaxy may appear as two different \textit{components} in the catalog while still belonging to the same physical \textit{source}.
Host galaxy identifications for detections in this catalog will be detailed in an upcoming work (Gordon et al., in prep).

The data used and the production of the VLASS epoch 1 \textit{Quick Look} component catalog are described in Section \ref{sec:catprod}.
The rapid CLEANing of the \textit{Quick Look} images leads to limitations of the scientific usefulness of this catalog, and these are described in Section \ref{sec:dataquality}.
We match our catalog to $\nu \sim 1.4\,$GHz and $\nu \sim 150\,$MHz data from FIRST and LoTSS respectively in Section \ref{sec:specidx}, and present the resultant spectral index and spectral curvature distributions.
In Section \ref{sec:vlassdnds} we present the VLASS source counts.
We use our catalog to explore the size distribution of VLASS components in Section \ref{sec:angsizes}.
This is complemented by an analysis of the VLASS two-point correlation function, and by exploring the impact of a factor of $\sim 2$ improvement in angular resolution over FIRST.
A summary of this paper is given in Section \ref{sec:summary}.
Throughout this work we adopt the convention $S_{\nu}\propto \nu^{\alpha}$ for radio spectral index measurements, and assume a flat $\Lambda$CDM cosmology with $h=0.7$, $H_{0} = 100h\,\text{km}\,\text{s}^{-1}\,\text{Mpc}^{-1}$, $\Omega_{\text{M}}=0.3$, and $\Omega_{\Lambda}=0.7$.

\section{catalog Production} 
\label{sec:catprod}

\subsection{VLASS Quick Look Images}


The first-epoch observations from VLASS are currently available as rapidly-processed \textit{Quick Look} images. 
In this paper we describe a catalog of radio components produced using these images, the availability of which was first announced in \citet{Gordon2020}.
These \textit{Quick Look} images are produced by NRAO and made available within weeks of the observations being conducted. 
However, the expedited nature of the image processing limits their quality, and hence constrains their scientific usability. 
Furthermore this rapid-CLEAN process employed by NRAO was updated between epoch 1.1 and 1.2 resulting in significant differences in image quality between the two sub-epochs \footnote{$49.9$ and $50.1\,\%$ of the imaging was performed in epoch 1.1 and 1.2 respectively.}. 
The epoch 1 \textit{Quick Look} images are described in full in  \citet{Memo13}, but we highlight here the key issues known in advance:
\begin{itemize}
    \item \textbf{Flux calibration} - The flux density values in the \textit{Quick Look} images are on average systematically low by $\sim 10\%$.
    Furthermore, \citet{Memo13} show that there is a substantial scatter, $\sim 15\,\%$, in the ratio of flux densities measured by VLASS and pointed observations of calibrator targets.
    
    \item \textbf{Astrometry} - The positional accuracy of the VLASS \textit{Quick Look} imaging is limited to $\sim1''$.
    This improves to $0''.5$ at declination, $\text{DEC,} > -20^{o}$.
    
    \item \textbf{Ghosts} - These appear offset by multiples of $\sim3'$ in right ascension, RA, east of bright sources. 
    The \textit{Quick Look} image production pipeline is designed to flag and remove these but some faint ghosts may still remain. 
    Moreover the removal of ghosts was refined between epoch 1.1 and 1.2 meaning that the earlier images are more likely to contain residual faint ghosts.
\end{itemize}
The impact of these reliability issues on our component catalog are further detailed in Section \ref{sec:dataquality}.

The full epoch 1 \textit{Quick Look} image set consists of $35,285$ one square degree \textit{Quick Look} images (hereafter referred to as subtiles) across 899 $\sim 40\,\text{deg}^{2}$ VLASS observing tiles \citep{Memo7}.
The catalog based on these data and used in this paper has been made available to the community via \url{https://cirada.ca/catalogues} \citep{Gordon2020}, and its production is described below.
An extensive User Guide is also available for download with the catalog, and we recommend reading this prior to accessing the data.


\subsection{Component Extraction and Duplicate Finding}

For each of the $35,285$ subtiles in VLASS, the source extraction code Python Blob Detection and Source Finding \citep[PyBDSF,][]{Mohan2015} was run in \textit{`srl'} mode - meaning that a list of components is provided rather than a list of individual Gaussian fits. 
The method adopted by PyBDSF is to detect flux islands at $3\sigma$ above the image mean (where $\sigma$ is the local rms), and then to fit components composed of one or more Gaussians within those islands with a peak brightness at $5\sigma$ above the image mean. 
For the most part the default PyBDSF parameters are used, with two exceptions:
\begin{enumerate}
    \item The sliding box used by PyBDSF to calculate the rms is set at a fixed size of $200\times200\,$px with a slide step of $50\,$px.
    Explicitly, the argument \text{\textit{rms\_box} $=\{200,50\}$} is called when running PyBDSF.
    The pixels in the subtiles are $1''$ square.
    
    \item For completeness PyBDSF is set to output detected islands of flux density to which it did not fit components in addition to the list of fitted components it finds.
    This is achieved by running PyBDSF with the argument \textit{inc\_empty} $=$True, and such detections are included and identified in the catalog by \textit{S\_Code} $==$ `E'.
    These `empty islands' are generally faint, with $95\,\%$ of such detections having a peak brightness of  less than $ 2\,$mJy/beam.
\end{enumerate}
The resultant $35,285$ individual PyBDSF component tables for each subtile are then merged into a single table. 
As there is a $\sim 2'$ overlap region between the images the catalog contains duplicate entries that refer to the same component. 
Based on the beam size and the positional uncertainty associated with the images, we use a $2''$ search radius around each component to identify duplicates. 
Where component duplicates are identified, preference is given to the component with highest ratio of peak brightness to local rms. 
All duplicates are retained within the catalog but are flagged as such (0=unique component; 1=duplicate component that is the preferred version of this component; 2=duplicate component that is not the preferred version of this component). 
To select a duplicate free list of components only components with \textit{Duplicate\_flag} $<2$ are used for the statistics presented in this work.

\subsection{Reliability of Detections}
\label{ssec:qualflags}

\begin{figure}
    \centering
    \subfigure{\includegraphics[width=0.49\columnwidth]{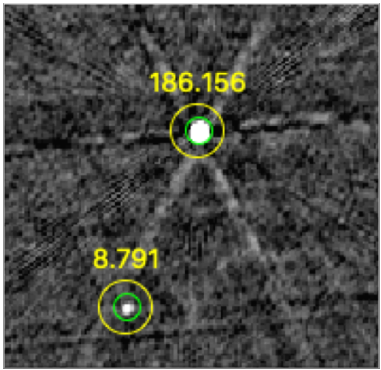}}
    \subfigure{\includegraphics[width=0.49\columnwidth]{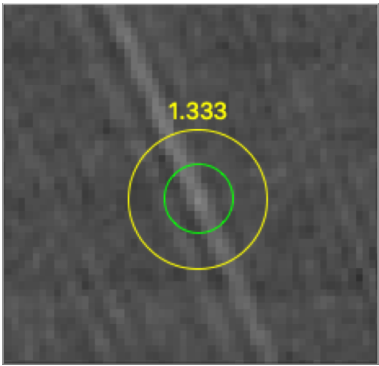}}
    \caption{The `peak-to-ring' metric is specifically designed to catch false-positive components that have been fit to artefacts, and more specifically sidelobes. 
    In these images the green ring shows the inner radius ($5''$) and the yellow ring the outer radius ($10''$) of the annulus in which flux density is compared to the component peak brightness.
    The left-hand panel shows two good component detections with high peak-to-ring values.
    The right hand plot shows a component erroneously fitted to a sidelobe.
    In this case the ratio of peak brightness to maximum flux in the annulus is only $1.33$, demonstrating that lower peak-to-ring values have diagnostic value in flagging false detections.}
    \label{fig:peak2ring}
\end{figure}

\begin{figure}
    \centering
    \subfigure{\includegraphics[width=0.45\columnwidth]{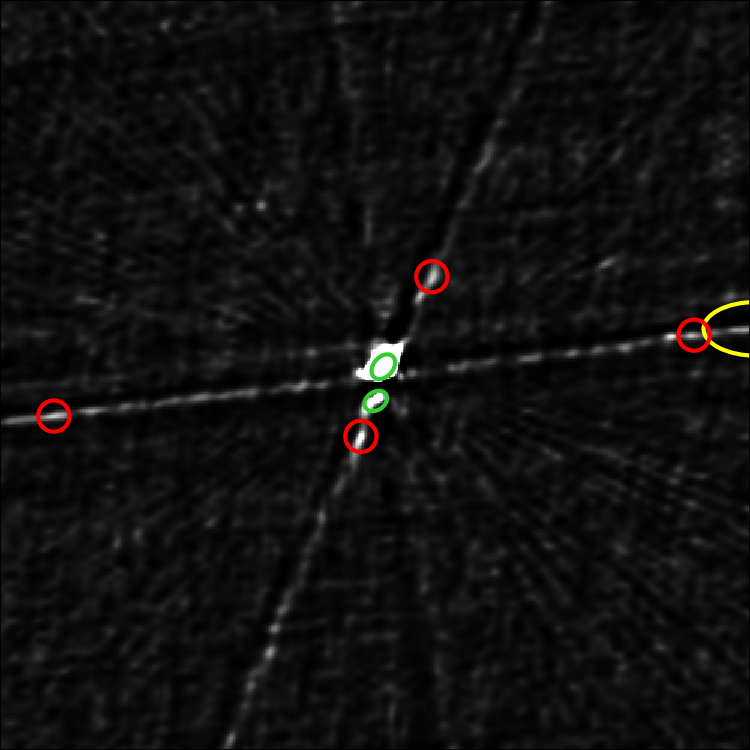}}
    \hspace{1mm}
    \subfigure{\includegraphics[width=0.45\columnwidth]{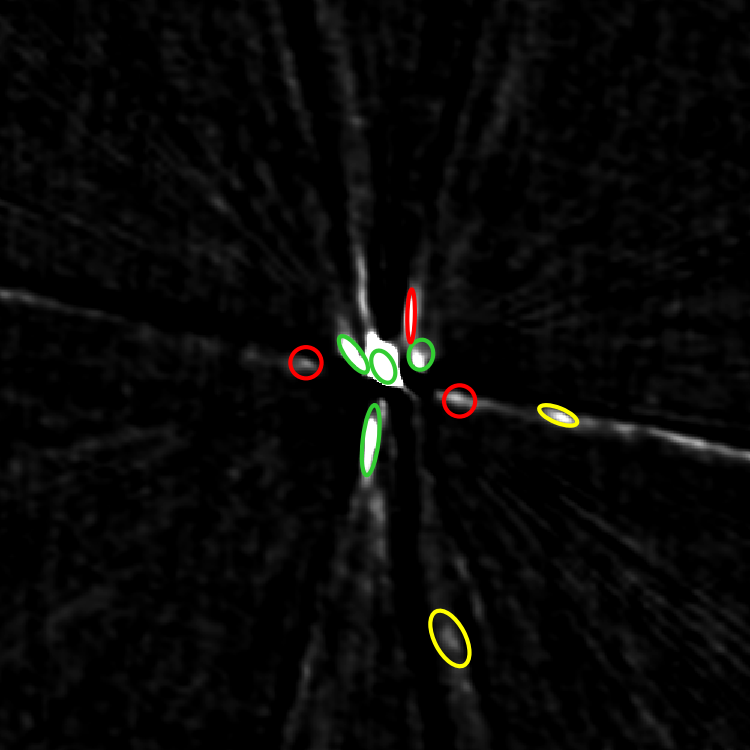}}
    \subfigure{\includegraphics[width=0.45\columnwidth]{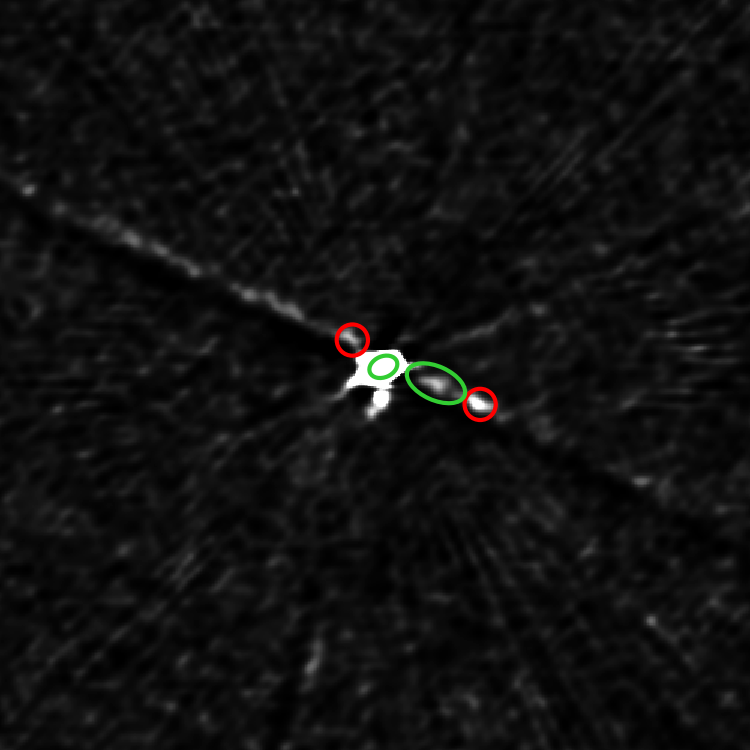}}
    \hspace{1mm}
    \subfigure{\includegraphics[width=0.45\columnwidth]{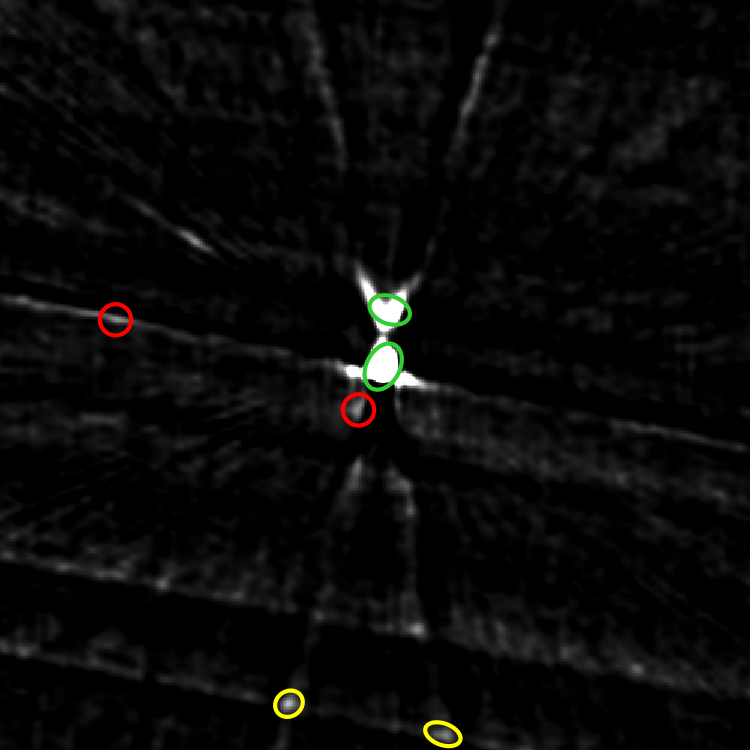}}
    \caption{Sample cutouts of $2' \times 2'$, centred on bright components demonstrating the sidelobes that the \textit{Quick Look} processing pipeline has failed to clean. 
    These artefacts can then be detected by PyBDSF as spurious components.
    Here green ellipses show components that our algorithm does not flag, yellow ellipses show components that our algorithm does flag, and red circles indicate where PyBDSF has detected a flux island but not fitted a component.
    The ellipse geometry is defined by the component parameters and enlarged by $50\%$ for clarity.
    }
    \label{fig:BScomponents}
\end{figure}

In addition to the basic measurements provided by PyBDSF, we wish to determine which detected components are real, as opposed to false positive detections.
A column, \textit{Quality\_flag}, is provided in our catalog to highlight our recommended components to use based on analysis of the raw output.
This takes account of three different parameters which are combined into a single integer flag value.
These are:
\begin{enumerate}
    \item The ratio of the total flux density and peak brightness measurements. 
    Given the reliability issues associated with the \textit{Quick Look} images, any component where the total flux is less than the peak brightness should be considered to be a possibly dubious measurement.
    Thus, we flag component with $S_{\text{peak}} > S_{\text{total}}$ by setting \textit{Quality\_flag} $=$ \textit{Quality\_flag}$+4$.
    
    \item The signal-to-noise (S/N) of the component.
    We flag components with a peak brightness lower than 5 times the local rms and set \text{\textit{Quality\_flag} $=$ \textit{Quality\_flag}$+2$}.
    Nearly all components flagged here have $3 < \text{S/N} < 5$, however we note that $1,600$ components have $\text{S/N} < 3$.
    This is attributed to differences in the local rms from the rms image used to detect the islands, and the fitted island rms.
    
    \item The ratio of peak brightness to the maximum flux density in a ring centred on the component position and with inner and outer radii of $5''$ and $10''$ respectively. 
    This is specifically designed to identify potential sidelobe structures that have erroneously been fitted as components.
    Sidelobes should have a low value for this ratio compared to true components that are surrounded by just the background noise (see Figure \ref{fig:peak2ring}).
    A caveat here is that close-double radio sources, the detection of which should be a forte of VLASS, will also provide low values to this ratio, and could thus be inadvertently  be identified as sidelobe detections. 
    In light of this, this metric is only used to flag components that are more than $20''$ from another component and with \textit{Peak\_to\_ring}$\ <2$.
    Here we apply \textit{Quality\_flag} $=$ \textit{Quality\_flag}$+1$.
\end{enumerate}
The combination of these parameters gives a single flag with an integer value between $0$ and $7$. 
In Figure \ref{fig:BScomponents} we show examples of our quality flagging routine identifying spurious detections due to image artefacts around bright components.

\begin{figure}
    \centering
    \includegraphics[width=\columnwidth]{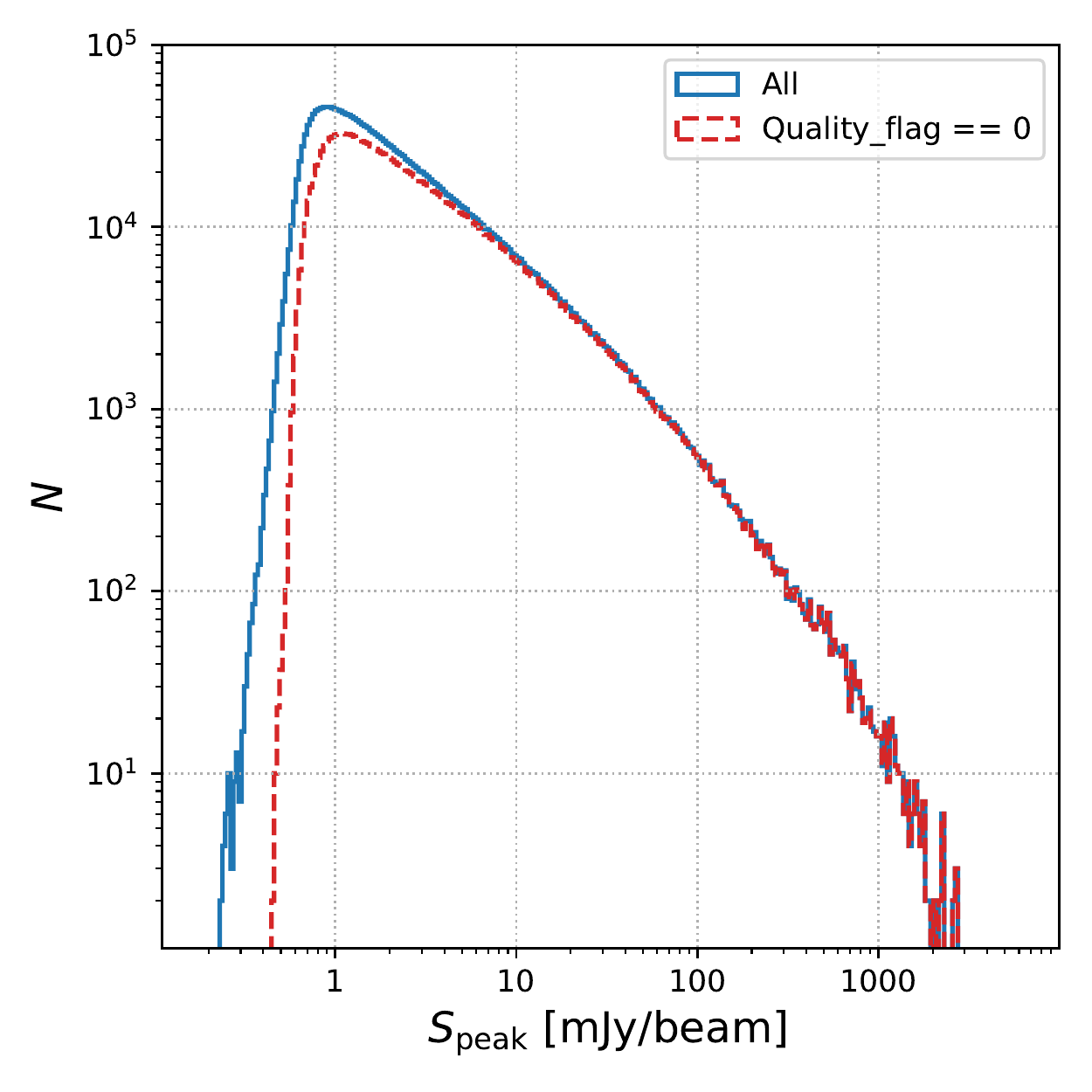}
    \caption{The distribution of peak brightness measurements from our catalog.
    The blue line shows the distribution for all fitted components, including those that do not have \textit{Quality\_flag}$ == 0$ (empty flux islands are not included here).
    The red line shows the flux distribution for only those components with \textit{Quality\_flag}$ == 0$. 
    Note the excess of low flux densities when using all components.}
    \label{fig:compsdist}
\end{figure}

In Figure \ref{fig:compsdist} we compare the peak brightness distributions for all components with components that have \textit{Quality\_flag} $==0$.
The distribution for components that have not been flagged shows a reduced number of faint detections.
Although this demonstrates that suspect detections are generally faint, note that the distribution for unflagged components deviates from the flagged component peak brightness distribution up to $\sim 5\,$mJy/beam.
Therefore, to select only the most reliable measurements, we recommend using \textit{Quality\_flag} $==0$, indicating that the component has not been flagged for any of the above quality issues, in addition to just applying simple brightness cuts to the data.

Whilst not explicitly provided by PyBDSF when set to provide component lists, we recover the pixel coordinates of components based on the image header information and provide them as additional information. 
Further information on the VLASS tile and subtile to which a detection belongs, and angular separation from the nearest other component (given in the column \textit{NN\_dist} for components with \textit{Duplicate\_flag} $<2$ and \textit{Quality\_flag} $==0$), are included.
Additionally, we provide the angular separations from the  nearest detections in the two previous NRAO wide-field continuum surveys, NVSS and FIRST, as the columns \textit{NVSS\_distance} and \textit{FIRST\_distance}\footnote{As FIRST only covers around $1/3$ of the VLASS footprint, \textit{FIRST\_distance} should only be considered useful within the FIRST footprint.} respectively. 
These latter two metrics may be useful to further assess the reliability of the detection (particularly for fainter sources), and for finding potential variable objects (in which case we urge caution on the part of the user and remind them of the systematic underestimation of fluxes in the VLASS \textit{Quick Look} imaging).
The resultant component catalog consists of $3,381,277$ rows where each row corresponds to a detected component or empty flux island. 
Of these, $1,692,158$ have \textit{Duplicate\_flag} $<2$ and \textit{Quality\_flag} $==0$. 

\section{Data Quality}
\label{sec:dataquality}


Here we detail the issues with data both known about from \citet{Memo13} and discovered during our quality assurance testing.
Whilst the major known issues are described here this list is unlikely to be comprehensive\footnote{We encourage users to report any additional issues they encounter with this data via the catalog feedback portal at \url{https://cirada.ca/catalogues}.}.

\begin{figure*}
    \centering
    \includegraphics[width=2\columnwidth]{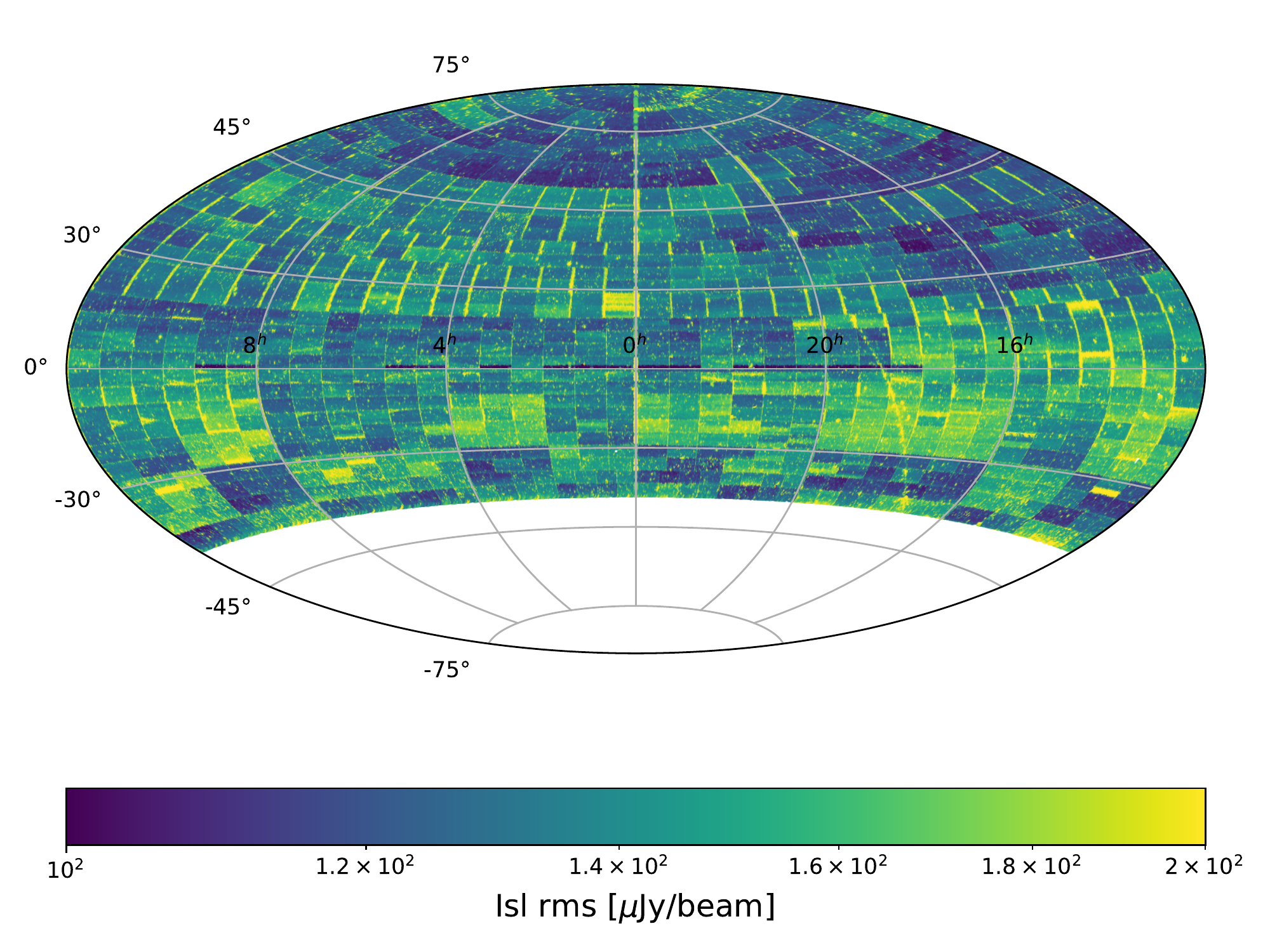}
    \caption{Aitoff projection in equatorial (J2000) coordinates of the rms values found around all VLASS components.
    For clarity only $100 < \text{Isl\_rms} < 200\,\mu$Jy/beam are shown and on a $\log_{10}$ scale.
    Additionally this is limited to components with \textit{Quality\_flag} $==0$ and \textit{Duplicate\_flag} $<2$.
    From this plot it is clear that the depth of the \textit{Quick Look} imaging is not homogeneous.
    Key features in this plot include the Galactic plane at low Galactic longitudes, the checker-board pattern at southern latitudes, high noise at the East-West boundaries of tiles, and a region of high noise at $\text{DEC} \sim +85^{o}$ in the Western hemisphere ($12 < \text{RA} < 24\,$hr).}
    \label{fig:aitoffrms}
\end{figure*}

\subsection{Noise Variation Between Images}



Between epoch 1.1 and 1.2 the VLASS \textit{Quick Look} image pipeline was upgraded to include automatic as opposed to manual flagging of science data.
Whilst reducing the human workload, this automatic routine increased the amount of flagging, resulting in a reduced bandwidth and thus noisier data (Mark Lacy, private communication).
It follows that these differences in the image processing between the two data sets will impact upon the sample homogeneity.
The median rms for images from epoch 1.1 is $128\,\mu$Jy/beam compared to $145\,\mu$Jy/beam for epoch 1.2, indicating the impact of the change of image processing methods on the data.
In Figure \ref{fig:aitoffrms} we show an Aitoff projection of the local rms around components in our catalog.
Inspecting this plot shows several curious features highlighting the lack of homogeneity of the VLASS \textit{Quick Look} imaging. 
In particular, users of this catalog should be aware of the following:
\begin{enumerate}
    \item At southern declinations a `checker-board' pattern of rms is apparent.
    This clear patterning is indicative of variation between VLASS tiles, and moreover of the difference between the two sub-epochs \citep[see also the VLASS tiling pattern described in][]{Memo7}.
    
    \item At Northern latitudes in particular, high noise levels are visible at the East-West boundaries of VLASS tiles.
    The increase in noise at specifically the East-West boundary is likely attributable to the fact that some tile edges were flagged to avoid unreliable fluxes associated with online software bugs related to the ghosts \citep{Memo13}.
    
    \item There is a region of high noise that stands out at $\text{DEC} \sim +85^{o}$ and $12\,\text{hr} < \text{RA} < 24\,$hr.
    
    \item The Galactic plane is clearly visible as a region of enhanced noise, albeit only at very low Galactic latitudes ($|b|\lesssim 0.5^{o}$) and for a relatively narrow range of Galactic longitudes ($350^{o} \lesssim l \lesssim 50^{o}$).
    
    \item There are several strips at $0^{o} < \text{DEC} < +1^{o}$ which have lower noise ($\sim 100\,\mu$Jy/beam).
    This is the result of accidentally observing these regions twice during survey operations and using both observations in the production of the \textit{Quick Look} images.
    This deeper VLASS region has a total area of $\sim 160\,\text{deg}^{2}$.
     
\end{enumerate}


\subsection{Reliability of Flux Density Measurements}
\label{ssec:fluxreliability}

From  \citet{Memo13} we know the flux densities in the VLASS \textit{Quick Look} imaging are systematically underestimated.
Based on comparisons with $>50$ VLA calibrator sources, \citet{Memo13} estimate that the peak brightness in the VLASS \textit{Quick Look} images are underestimated by $\sim15\,\%$, and the total flux density of components is underestimated by  $\sim10\,$\%. \citet{Memo13} report a scatter of about $8\,\%$ in these measurements.
In this Section we compare the flux density measurements in our catalog to existing survey data so as to obtain our own estimate of the VLASS flux density reliability.

\begin{figure}
    \centering
    \subfigure{\includegraphics[width=0.49\textwidth]{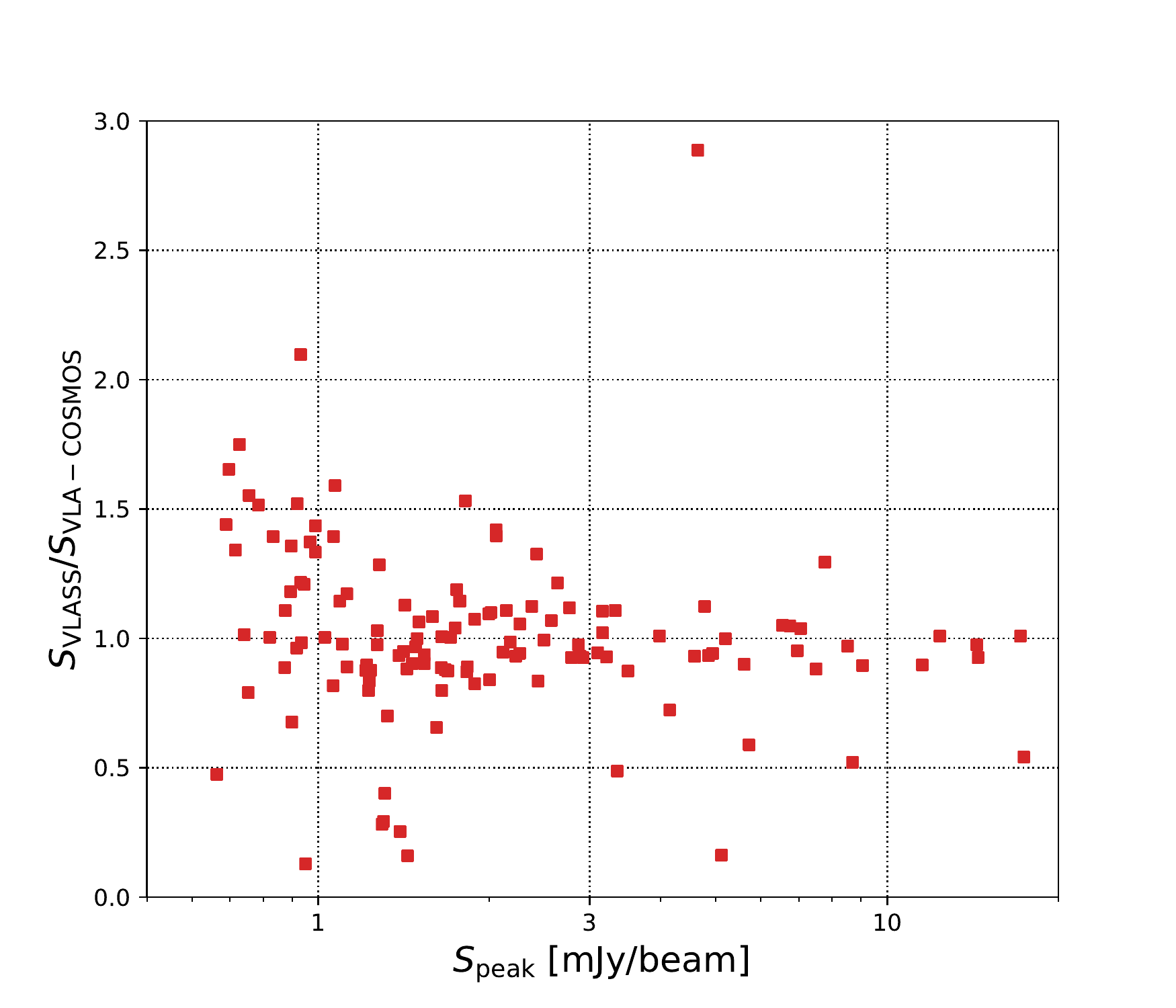}}
    \caption{Comparison of VLASS flux measurements to the 3GHz VLA-COSMOS measurements for the 131 matched components in the 2 square degrees of the COSMOS field.
    }
    \label{fig:vlacosflux}
\end{figure}

\subsubsection{Direct comparisons with other 3~GHz data}
\label{sssec:vlacoscomparison}

One approach to measure the impact of the potential flux density underestimation of the VLASS \textit{Quick Look} images on our catalog is to directly compare our measurements to existing $3\,$GHz flux densities for components in our catalog.
However, previous blind continuum observations at this frequency have come from deep, narrow-field surveys \citep{Vernstrom2016, Smolcic2017}, limiting the overlap in detected components with VLASS.
Of the available $3\,$GHz catalogs, the VLA-COSMOS $3\,$GHz Large project \citep[hereafter referred to simply as VLA-COSMOS,][]{Smolcic2017}, covering the two square degrees of the COSMOS field, presents the largest overlap with VLASS in terms of sensitivity and sky coverage.
Our VLASS catalog contains $131$ components with a VLA-COSMOS detection within $2''.5$ ($126$ of these are separated by less than $1''$).

The small number of overlapping components between VLASS and VLA-COSMOS limits our ability to perform a robust flux calibration of our catalog via a direct comparison of the independent flux density measurements at $3\,$GHz.
Nonetheless, we report here our findings in comparing the measurements from both catalogs.
Given the already small sample size, for this comparison we do not account for the difference in resolution between VLA-COSMOS ($0''.75$) and VLASS.
In Figure \ref{fig:vlacosflux} we show the ratio of VLASS to VLA-COSMOS flux density measurements for the $131$ matched components as a function of peak brightness in VLASS.
This sample contains $33$ components with a peak VLASS brightness greater than $3\,$mJy/beam, and for these the median value of $S_{\text{VLASS}}/S_{\text{VLA-COSMOS}}$ is $0.95$ with a standard error of $0.07$.
Below about $3\,$mJy/beam there is substantial scatter in the ratio of measured flux densities between VLASS and VLA-COSMOS.

\subsubsection{Comparisons with multi-band radio data}
\label{sssec:fluxcal}


\renewcommand\arraystretch{1.3}
\begin{table*}
    \centering
    \caption{Parameters used when matching comparison surveys with VLASS for flux calibration.
    The frequency ($\nu$) of the survey in MHz, the minimum angular distance to the nearest neighbor for the matched VLASS component in arcseconds, the minimum flux density of components from that survey used in mJy are listed below for each survey.
    The number of matched components for which the VLASS to comparison survey flux density ratios are used to calibrate the VLASS flux are given in the column $N$}
    \begin{tabular}{ l | c c c c c}
         \textbf{Survey} & \textbf{$\nu$} & \textbf{Isolation radius} & \textbf{$\Psi_{\text{max}}$} & \textbf{$S_{\text{min}}$} & \textbf{$N$}\\
          & \textbf{[MHz]} & \textbf{[arcsec]} & \textbf{[arcsec]} & \textbf{[mJy]} &  \\
         \hline 
         TGSS & $150$ & $14$ & $30$ & $60$ & $806$\\
         WENSS & $330$ & $54$ & $0$ & $20$ & $1,636$\\
         SUMSS & $843$ & $43$ & $10$ & $10$ & $1,747$\\
         FIRST & $1,400$ & $2.7$ & $1$ & $2$ & $6,189$\\
         VLASS & $3,000$ & -- & $0.5$ & $3$ & --\\
    \end{tabular}
    \label{tab:compsurveys}
\end{table*}

Given the limited available $\nu \sim 3\,$GHz data with which to compare our catalog, we adopt a method similar to that used by \citet{Sabater2021} in order to estimate the reliability of our flux density measurements.
This approach determines the distribution of flux density ratios for matched components between the survey being calibrated and a reference survey at a number of different frequencies.
A power law can then be fit to the average flux density ratio as a function of frequency, with the intercept at the frequency of the survey being calibrated used to determine the flux density accuracy.
As the VLASS footprint covers $\sim 80\,\%$ of the sky there are a number of overlapping wide-field continuum surveys that can be used in this way.

We compare our data with TGSS at $150\,$MHz, the Westerbork Northern Sky Survey \cite[WENSS,][]{Rengelink1997} at $330\,$MHz, the Sydney University Molonglo Sky Survey \citep[SUMSS,][]{Bock1999, Mauch2003} at $840\,$MHz and FIRST at $1.4\,$GHz.
As all these surveys have different angular resolutions and effective depths\footnote{defined as the noise level when accounting for observing frequency assuming a typical spectral index of $-0.7$; higher (lower) effective noise levels imply shallower (deeper) surveys} care must be taken when matching components.
For instance, WENSS has an angular resolution of $54''\,\csc$(DEC) and a single WENSS component could easily be observed as multiple components in VLASS.
Therefore, to qualify as a matched components  we require the distance to the nearest neighboring VLASS component to be greater than half the angular resolution of the comparison survey, using the \textit{isolation radius} given in Table \ref{tab:compsurveys}.
As the elliptical beam geometry varies with declination for TGSS, WENSS and SUMSS, the poorest possible resolution in the overlap with VLASS is assumed when defining this isolation limit for these surveys.

Only unresolved or barely resolved components are matched to one another in this calibration. 
For VLASS, we consider components with a deconvolved angular size, $\Psi$, of less than $0''.5$ to be suitable here.
The maximum angular sizes of components we use from the comparison surveys are listed in Table \ref{tab:compsurveys}, and for FIRST and SUMSS these are based on the cataloged deconvolved sizes.
The WENSS catalog provides the Gaussian fitted sizes (i.e. including the beam size) for resolved components, and lists a `zero' size where the ratio of peak and integrated flux density is indicative that the component is unresolved \citep{Rengelink1997}. 
It is these unresolved WENSS components we include in our analysis.
TGSS only provides a fitted component size and does not highlight likely unresolved components, so we only consider TGSS components with a fitted angular size smaller than $30''$ in this analysis.

The appropriate samples of isolated VLASS components are then cross matched with their respective comparison surveys.
As TGSS, WENSS and SUMSS have substantially poorer angular resolutions than VLASS we use a search radius of $5''$ when cross matching these surveys with VLASS.
For FIRST, which has an angular resolution ($5''.4$) that is more comparable to that of VLASS, a search radius of $2''.5$ is used. 

\begin{figure}
    \centering
    \subfigure{\includegraphics[trim=0 0 0 0, clip, width=0.46\textwidth]{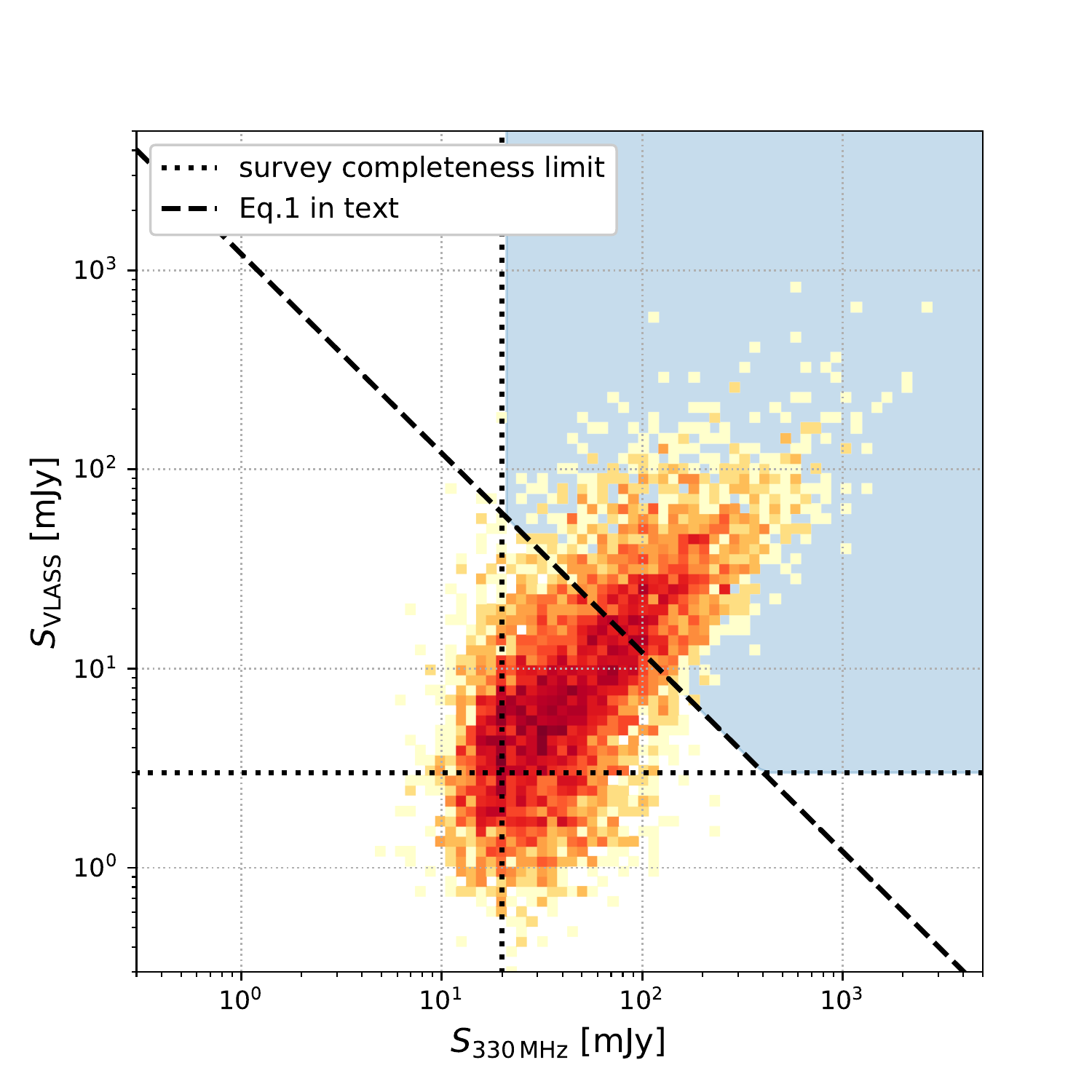}}
    \subfigure{\includegraphics[trim=0 0 0 15mm, clip, width=0.46\textwidth]{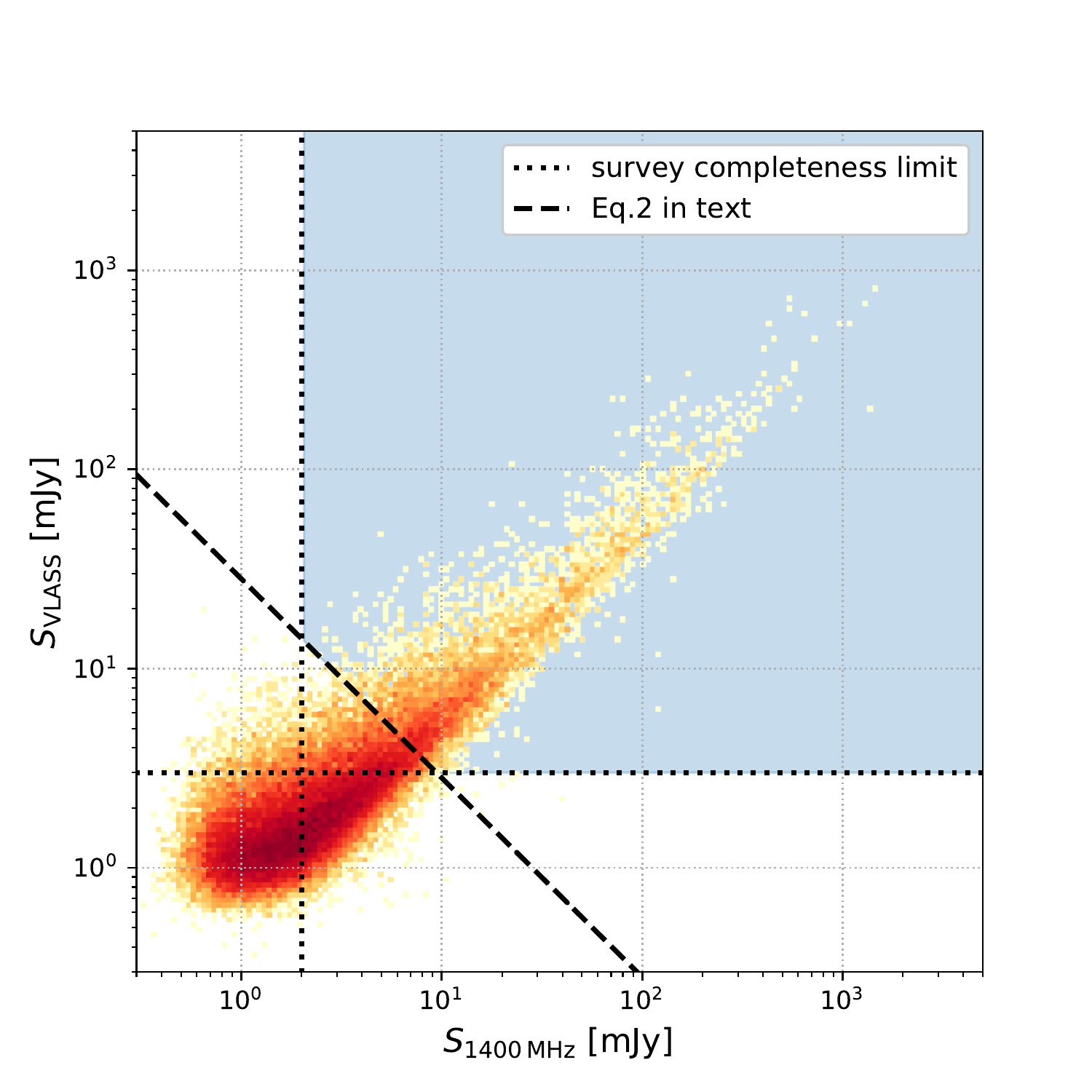}}
    \caption{Examples of VLASS flux densities compared to two of the comparison surveys.
    The upper panel shows the comparison with a shallower survey,  WENSS ($\nu\sim 330\,$MHz), and the lower panel shows the comparison with the deeper survey, FIRST ($\nu \sim 1,400\,$MHz).
    The red/yellow density plots show the flux densities in VLASS and the comparison survey for \textit{all} cross matched compact components.
    The black dashed lines show the minimum flux density used as a completeness limit for each survey, while the solid black line shows the cutoff we apply to prevent the different effective survey depths biasing our flux calibration.
    For WENSS this is based on Equation \ref{eq:minvlassflux_scomp}, and for FIRST this is based on Equation \ref{eq:mincompflux_scomp}.
    The light blue shaded region above and to the right of the three black lines highlight the components selected for use in our VLASS flux calibration.}
    \label{fig:fluxcompeg}
\end{figure}

All of the comparison surveys are at a lower frequency than VLASS; comparing to a survey with a greater effective depth (e.g., FIRST) will be biased by components with flat- or inverted-spectrum sources.
Comparisons with lower effective depth surveys (e.g. WENSS) will be biased towards components with very steep negative spectral indices.
This effect can be seen in the curvature at the low ends of the flux density comparisons for VLASS, WENSS and FIRST shown in Figure \ref{fig:fluxcompeg}.
To eliminate this bias, we  select only components with a spectral index, $\alpha$, in the range $-1.5 < \alpha < +0.5$, that would be detectable in both VLASS and the comparison survey.
Explicitly, this is achieved by satisfying :
\begin{equation}
	\label{eq:minvlassflux_scomp}
	S_{\text{VLASS}} > \frac{S_{\text{min, comparison}}^{2}}{S_{\text{comparison}}} \times \Bigg(\frac{3\,\text{GHz}}{\nu_{\text{comparison}}}\Bigg)^{+0.5},
\end{equation}
when the comparison survey is effectively shallower (TGSS, WENSS and SUMSS), and by
\begin{equation}
	\label{eq:mincompflux_scomp}
	S_{\text{VLASS}} > \frac{(3\,\text{mJy})^{2}}{S_{\text{comparison}}} \times \Bigg(\frac{\nu_{\text{comparison}}}{3\,\text{GHz}}\Bigg)^{-1.5},
\end{equation}
when the comparison survey is effectively deeper (FIRST).
$S_{\text{VLASS}}$ is the VLASS flux density of the component,
$S_{\text{comparison}}$ is the flux density of the component in the comparison survey, 
$S_{\text{min, comparison}}$ is the completeness limit of the comparison survey (see Table \ref{tab:compsurveys}),
$\nu_{\text{comparison}}$ is the comparison survey frequency,
and the completeness limit of VLASS is taken to be $3\,$mJy.
This selection is shown by the light blue shaded regions in Figure \ref{fig:fluxcompeg} for WENSS and FIRST.

\begin{figure}
    \centering
    \includegraphics[trim=0 0 0 0, clip, width=0.49\textwidth]{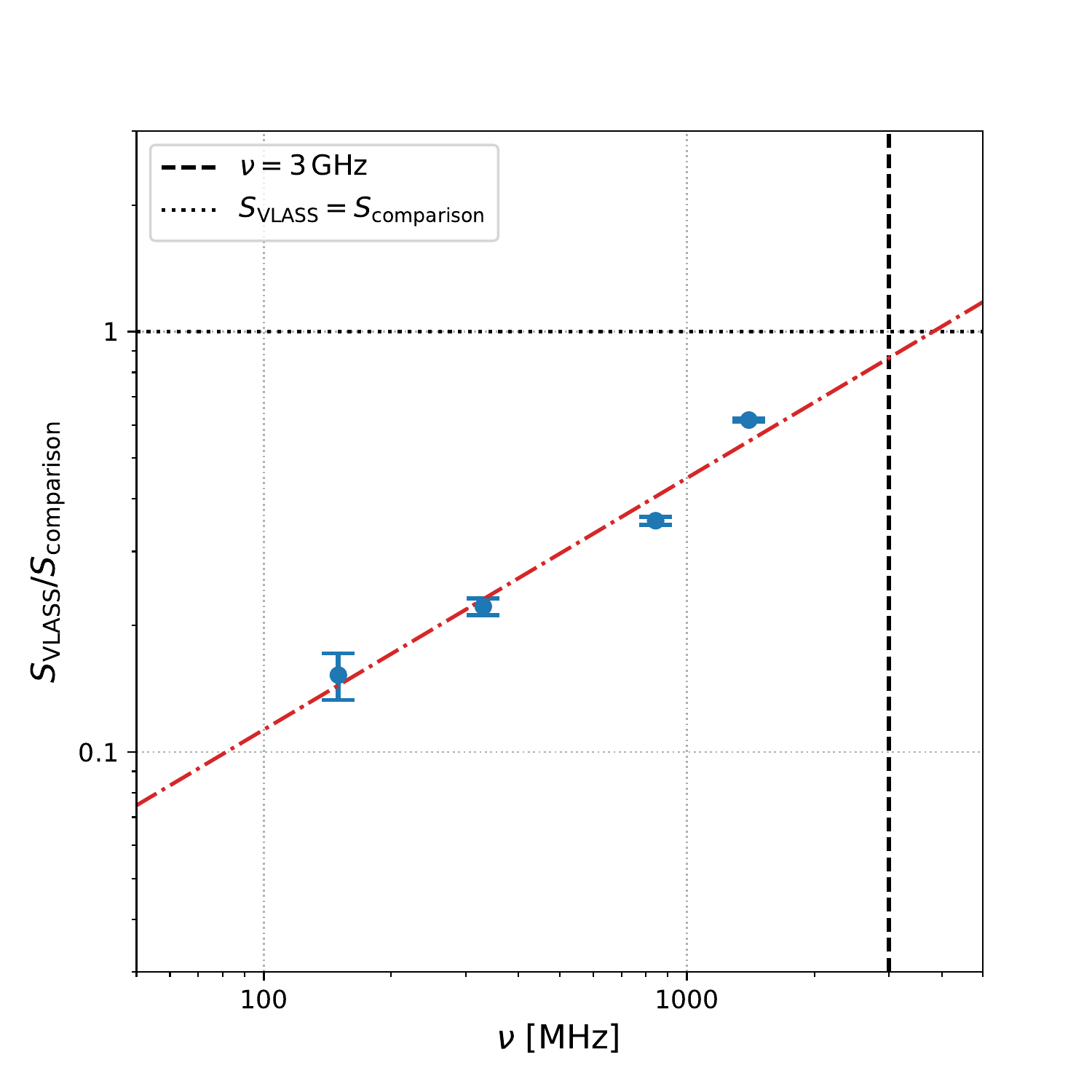}
    \caption{The median ratio of VLASS to comparison survey flux density for cross matched components as a function of comparison survey frequency.
    Error bars represent the standard error.
    Four comparison surveys are used here: TGSS at $150\,$MHz, WENSS at $330\,$MHz, SUMSS at $840\,$MHz and FIRST at $1,400\,$MHz.
    The red dash-dotted line shows the best fit to the data, a power law with a slope of $0.6$, from which we extrapolate $S_{\text{VLASS}}/S_{3\,\text{GHz}}=0.87$.
    The black dotted line shows unity for the ratio of VLASS to comparison flux densities, and the black dashed line marks $\nu = 3\,$GHz.}
    \label{fig:fluxcal}
\end{figure}

For the selected components we determine the median ratio of flux densities between VLASS and the comparison survey, and these are shown as a function of frequency in Figure \ref{fig:fluxcal}.
We fit a power law to these data points with a slope of $0.6$ in order to extrapolate the expected flux density ratio between VLASS and benchmark $3\,$GHz measurements, and thus quantify the typical underestimation of flux density measurements in our catalog.
From this we determine $S_\text{VLASS} = 0.87\, S_{3\,\text{GHz}}$.
To compensate for this typical underestimate, we scale all the VLASS flux density measurements by $1/0.87$ throughout the rest of this work. 
This correction is not applied to our catalog and anyone making use of this data is advised to apply the flux density correction themselves.

\subsection{Astrometry}

\begin{figure}
    \centering
    \subfigure{\includegraphics[width=0.49\textwidth]{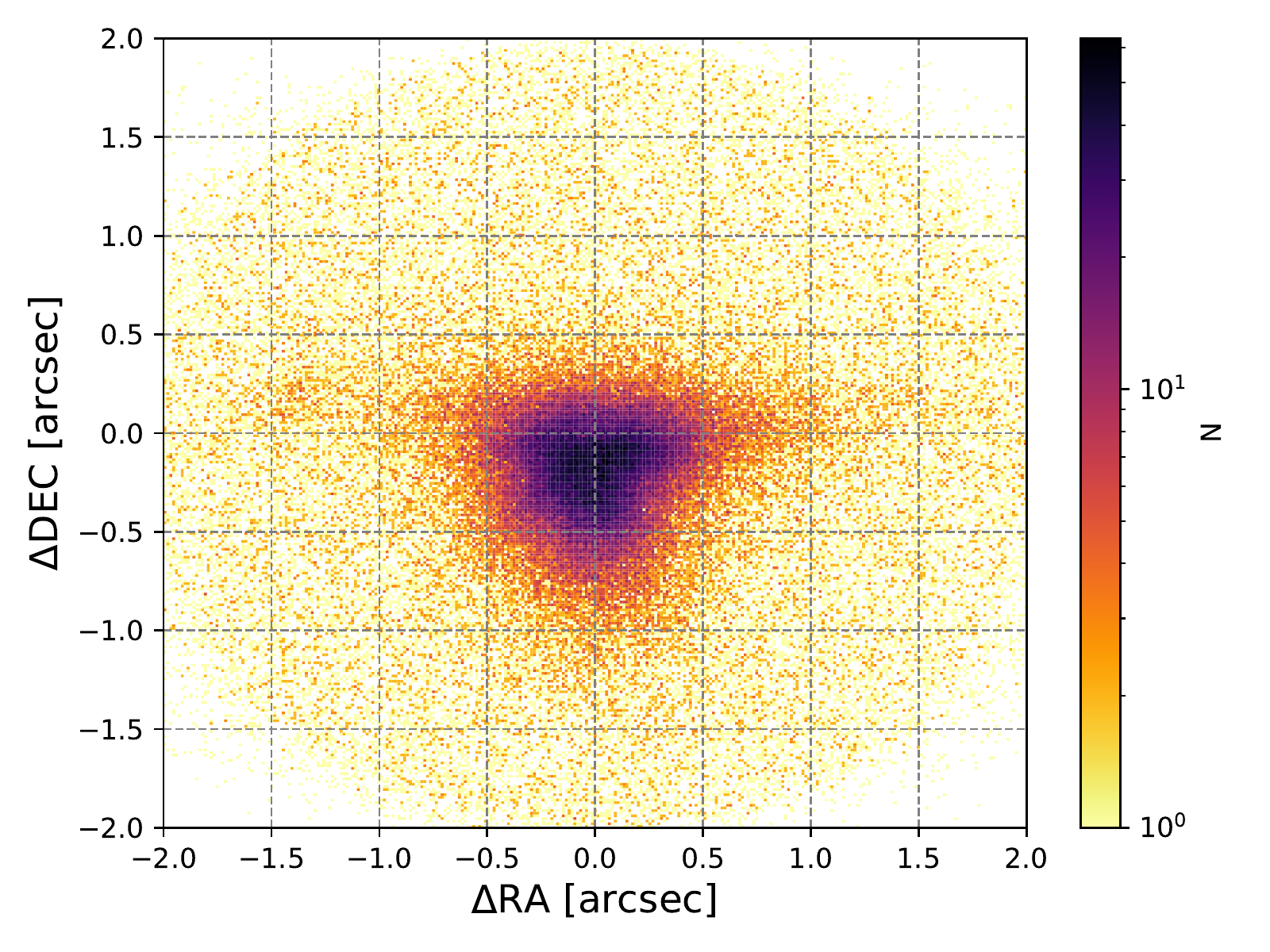}}
    \subfigure{\includegraphics[width=0.49\textwidth]{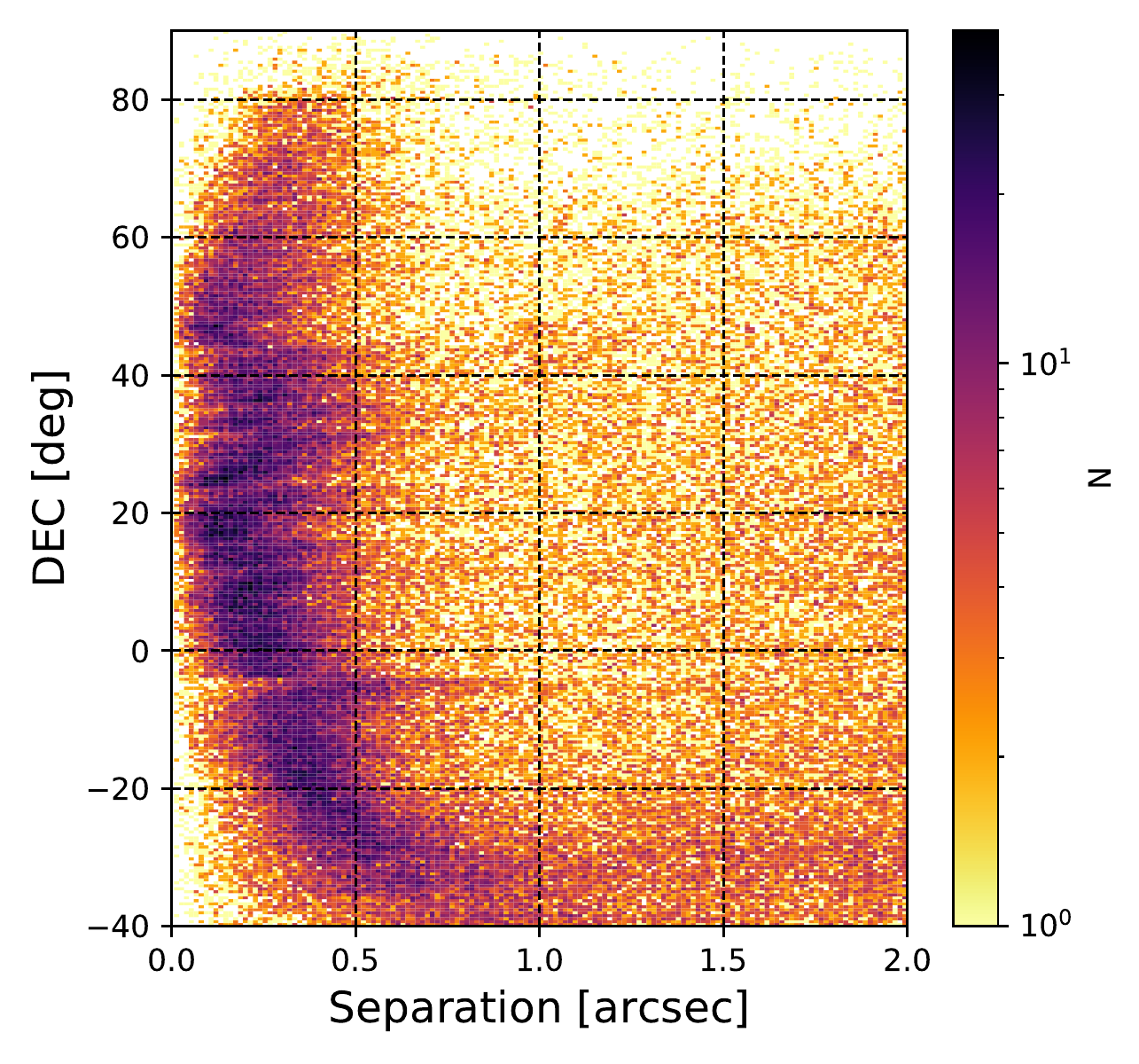}}
    \caption{The positional offsets of Gaia and VLASS sources separated by less than $2''$.
    The upper panel shows the offset in terms of RA and DEC. Here $\Delta x = x_{\text{Gaia}} - x_{\text{VLASS}}$, where $x = $ RA, Dec.
    While the peak of $\Delta$RA is close to 0, $\Delta$Dec peaks at around $-0''.25$ showing the VLASS \textit{Quick Look} imaging astrometric errors to be dominated by offsets in declination.
    The lower panel shows the separation, $\sqrt{\Delta\text{RA}^{2} + \Delta\text{Dec}^{2}}$, between VLASS and Gaia sources as a function of declination.
    At $\text{DEC} < -20^{o}$ the peak of the positional offset tends to higher separations indicating poorer astrometry at these declinations.}
    \label{fig:gaiaradec}
\end{figure}

Early investigations by NRAO into the epoch 1.1 \textit{Quick Look} imaging showed issues with respect to the astrometric accuracy of the \textit{Quick Look} images \citep{Memo13}.
Comparisons with Gaia DR2 \citep{Gaia2016, Gaia2018} showed the VLASS component positional accuracy to be limited to $\sim1''$, improving to $\sim0''.5$ at $\text{DEC} >-20^{o}$.
These limitations to the VLASS astrometry are attributable to two factors.
Firstly, the rapid processing technique employed fails to account for the $w$-term in the interferometer equation \citep{Smirnov2011a, Smirnov2011b}.
The impact of this is to distort the VLASS point spread function away from the phase center, resulting in positional offsets in detections.
Secondly, the pixels in the \textit{Quick Look} images are $1''$ compared to a typical VLASS beam size of $2''.5$.
Thus, the pixels in the \textit{Quick Look} do not adequately sample the beam, and this may contribute to the positional uncertainty.

Comparing our catalog with Gaia DR2 we confirm NRAO's findings in this regard.
In Figure \ref{fig:gaiaradec} we show the positional offset explicitly as $\Delta$RA and $\Delta$Dec, where $\Delta x = x_{\text{Gaia}} - x_{\text{VLASS}}$ for $x = $ RA, Dec.
The separation between VLASS components and Gaia sources is strongly peaked due to genuine associations.
If the astrometry for both surveys were perfect, one would expect the peak from genuine associations to occur at $\theta = 0''$, but we observe that the declination offset peaks at $\Delta\text{Dec} \approx -0''.25$. 
The offset in RA is minimal with a median of $\Delta\text{RA} \approx 0$.
In agreement with the findings in  \citet{Memo13}, we also confirm that the positional offset is worse at $\text{DEC} > +60^{o}$ and $\text{DEC} < 0^{o}$, with the separations between VLASS and Gaia sources mostly being $< 0''.5$ at $\text{DEC} > -20^{o}$ (see Figure \ref{fig:gaiaradec}).

\subsection{Data Selection}

In Section \ref{ssec:qualflags} we detailed the efforts we have undertaken to identify the spurious detections that are present in our VLASS \textit{Quick Look} catalog.
While a selection based on \textit{Quality\_flag} $== 0$ provides a sample containing the most reliable measurements, this may not be a complete sample, and thus will hinder a statistical characterisation of the survey.
In particular, the choice to exclude components where \textit{Peak\_flux}$ > $\textit{Total\_flux} will remove a substantial number of real detections that are unresolved by VLASS, in addition to removing low signal-to-noise false positive detections.

\begin{figure}
    \centering
    \includegraphics[width=\columnwidth]{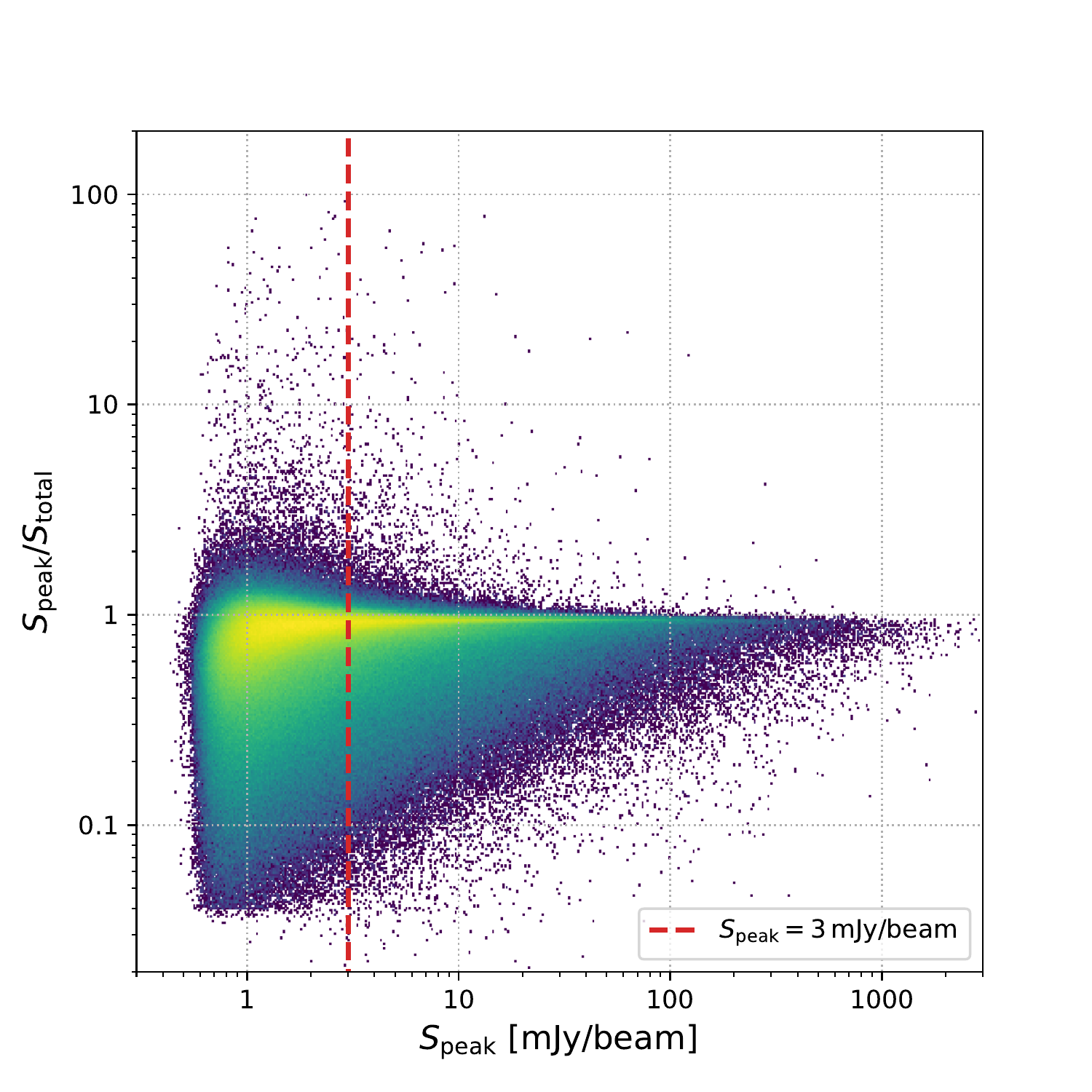}
    \caption{The ratio of peak brightness to total flux density as a function of peak brightness for the $\sim 1.9\times 10^{6}$ components in our VLASS catalog with \textit{Quality\_flag} $== (0|4)$, \textit{Duplicate\_flag} $ < 2$ and \textit{S\_Code} $\neq$ `E'.
    The red dashed line shows the brightness level of $3\,$mJy/beam used in this work to select components with the most reliable flux density measurements (see also Section \ref{ssec:fluxreliability}).
    The large population of components with $S_{\text{peak}}/S_{\text{total}} < 1$ represent well resolved VLASS components.
    Components with $S_{\text{peak}}/S_{\text{total}} > 1$ are unresolved components, and $\sim8\,\%$ of these are spurious in VLASS \textit{Quick Look} images.}
    \label{fig:peak2tot}
\end{figure}

To retain this population of unresolved components for the statistics presented in this paper, with the acceptance of source contamination by false positive detections at lower signal-to-noise, we select components from our catalog with \textit{Quality\_flag} $== (0|4)$, \textit{Duplicate\_flag} $ < 2$ and \text{\textit{S\_Code} $\neq$ `E'.}
This sample of $1,880,195$ components represents the $\sim1.7 \times 10^{6}$ selected by the original, higher fidelity criteria outlined in Section \ref{ssec:qualflags} (\textit{Duplicate\_flag} $<2$ and \textit{Quality\_flag} $==0$), 
with the addition of fitted components (i.e. not `empty islands') where \textit{Peak\_flux}$ > $\textit{Total\_flux}.
The distribution of the ratio of peak-to-total flux density as a function of peak brightness for these components is shown in \text{Figure \ref{fig:peak2tot}.}

In order to quantify the impact of potential spurious detections arising from the population of \text{\textit{Quality\_flag} $== 4$} components, i.e. those with $S_{\text{peak}} > S_{\text{total}}$ but otherwise unflagged in our catalog, we visually inspect a random sample of 100 of these components.
Here we find that $92_{-4}^{+2}\,\%$\footnote{errors are assumed to be binomial based on \citet{Cameron2011}} are clearly real detections.
As the \text{\textit{Quality\_flag} $== 4$} components account for $10\,\%$ of the sample this suggests a contamination level of $< 1\,\%$ spurious detections.
Moreover, for components with $S_{\text{peak}}> 3\,$mJy/beam used for most of the scientific analysis in this work (see Section \ref{sssec:vlacoscomparison}) only $6\,\%$ have \text{\textit{Quality\_flag} $== 4$}, suggesting a contamination level of $<0.5\,\%$ at this brightness level.
Throughout the rest of this work we use these $\sim1.9\times 10^{6}$ components with \textit{Quality\_flag} $== (0|4)$, \textit{Duplicate\_flag} $ < 2$ and \textit{S\_Code} $\neq$ `E' as our primary sample.
Many of the analyses presented are further restricted to the  $615,045$ components with $S_{\text{peak}}>3\,$mJy/beam where the reliability in the flux density measurements is best characterised (see Section \ref{ssec:fluxreliability}).


\section{Spectral Index and Curvature}
\label{sec:specidx}

\subsection{The VLASS/FIRST Spectral Index Distribution}
\label{ssec:spidx}

In the remainder of this work, we use the catalog of components from \textit{Quick Look} images to explore the scientific potential of VLASS.
The availability of wide field $\nu\sim 3\,$GHz observations at similar depths to previous surveys at $1.4\,$GHz lends itself to determining the distribution of spectral indices at these frequencies.
For this we cross-match VLASS and FIRST components within $2.''5$ of each other ($85\,\%$ are within $1''$), where the cross match radius is based on the typical VLASS beam size.
As in Section \ref{sssec:fluxcal} we require the VLASS component to be isolated in order to prevent the flux from the poorer resolution FIRST observations being split across multiple VLASS components (see also Section \ref{ssec:splitfirst}).
To enable further cross matching with LoTSS, that allows us to explore the spectral curvature of these object in Section \ref{ssec:speccurve}, we require the distance to the nearest neighbour in VLASS to be greater than $3''$ \textemdash\ half of the LoTSS beam size \citep[$6''$,][]{Shimwell2019}.
So as to limit issues relating to VLASS missing flux from extended low-surface brightness regions, only relatively small FIRST components, those with deconvolved sizes of $<10''$, are included in this sample.
We then determine the spectral index for $>500,000$ VLASS components using their VLASS and FIRST total flux density measurements. 

\begin{figure}
    \centering
    \subfigure{\includegraphics[trim=0 10mm 0 10mm, clip, width=1\columnwidth]{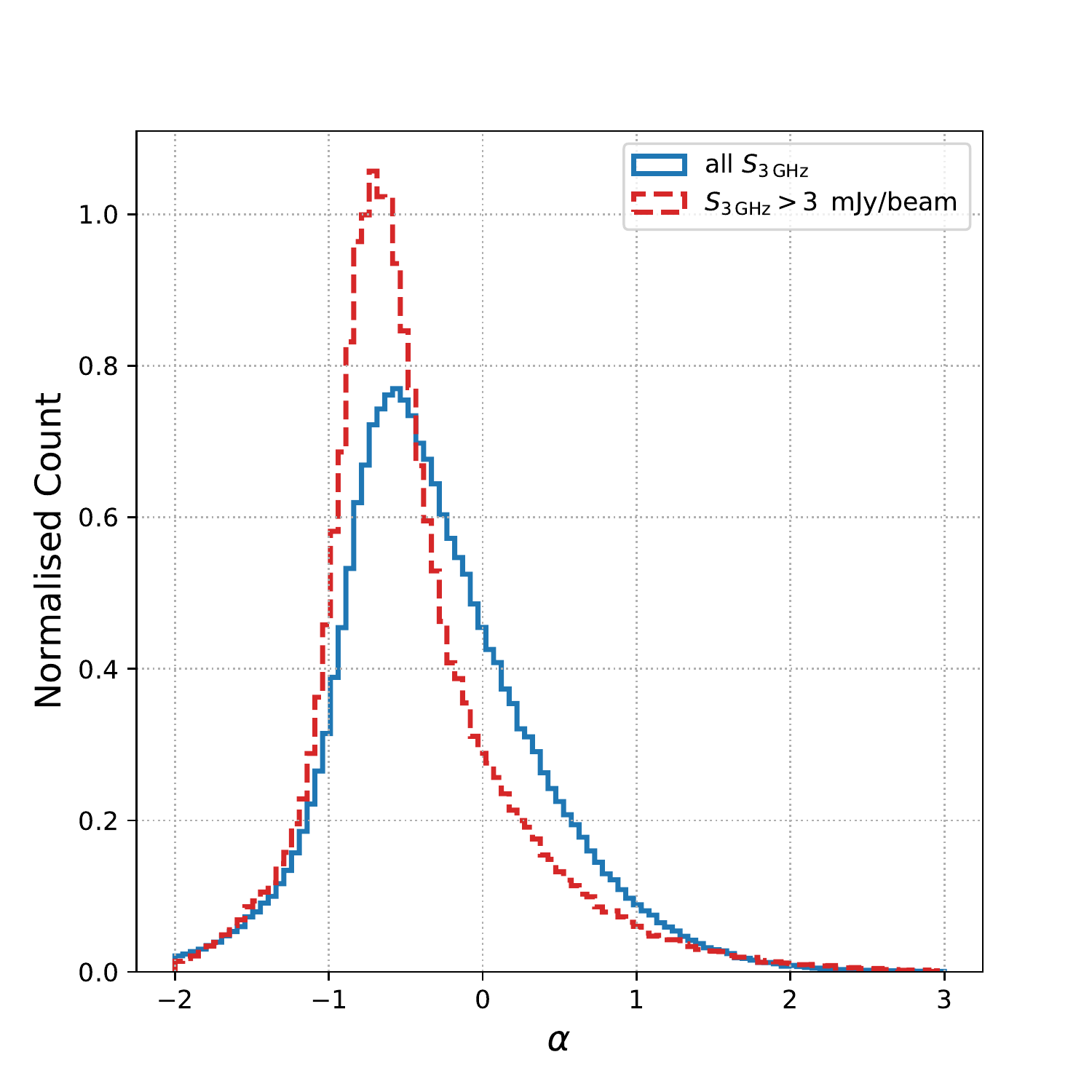}}
    \subfigure{\includegraphics[trim=0 5mm 0 12mm, clip, width=1\columnwidth]{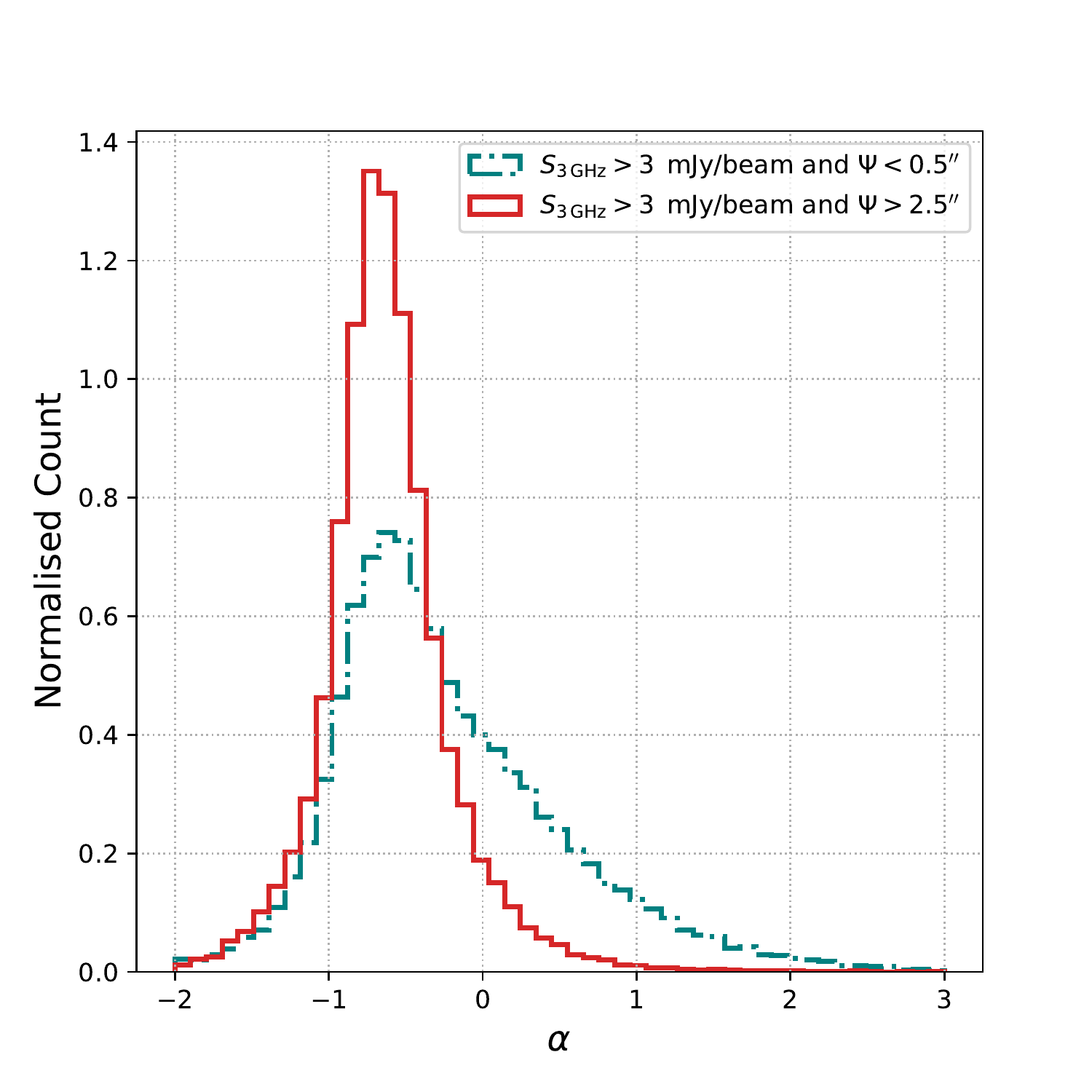}}
    \caption{Normalised distributions of the spectral index, $\alpha$, for VLASS components associated with FIRST components with deconvolved sizes of less than $10''$.
    The upper panel compares the spectral index distributions for VLASS components over the full range of detected flux densities (blue solid line) to those with $S_{3,\text{GHz}} > 3\,$mJy/beam (red dashed line).
    The lower panel compares components with deconvolved sizes greater than $2.5''$ (red solid line) to those smaller than $0.5''$ (blue/green dot-dashed line).
    We follow the convention that $S\propto \nu^{\alpha}$, and the spectral index is calculated using the total flux densities from the FIRST and VLASS catalogues.}
    \label{fig:specindex}
\end{figure}

In Figure \ref{fig:specindex} we show the two-point $1.4 - 3\,$GHz spectral index distribution, which peaks at $\alpha = -0.57$ and has a median of $\alpha=-0.39$, flatter than the typical value of $\alpha \sim -0.7$ one would expect based on previous works \citep[e.g.][]{Condon2002, Best2005, deGasperin2018}.
We showed in Figure \ref{fig:vlacosflux} that VLASS flux densities below $\sim 3\,$mJy are unreliable, and we thus limit our sample of matched VLASS and FIRST components to those with a peak brightness higher than $3\,$mJy/beam.
Doing this we find that the distribution peaks at $\alpha = -0.71$, in line with canonical values for the spectral indices of radio sources (see Figure \ref{fig:specindex}).
Setting an even higher limit on the minimum flux density of the sample of $S>10\,$mJy/beam, the peak of the distribution is located at $\alpha = -0.72$ (see Table \ref{tab:specindex}).

\renewcommand\arraystretch{1.3}
\begin{table*}
    \centering
    \caption{Peak and median values of the spectral index distributions for VLASS components matched to FIRST components with a deconvolved angular size, $\Psi$, less than $10''$. The number of components in each data set is given by $N$.}
    \begin{tabular}{l|c c c}
         \textbf{Component subset} & \textbf{$N$} & \textbf{Peak $\alpha$} & \textbf{Median $\alpha$}\\
         \hline 
         All & $533,739$ & $-0.57$ & $-0.39$\\
         $S_{3\,\text{GHz}}>3\,$mJy/beam & $182,638$ & $-0.71$ & $-0.58$\\
         $S_{3\,\text{GHz}}>10\,$mJy/beam & $57,181$ & $-0.72$ & $-0.65$\\
         $\Psi< 0.5''$ and $S_{3\,\text{GHz}}>3\,$mJy/beam & $33,279$ & $-0.66$ & $-0.38$\\
         $\Psi> 2.5''$ and $S_{3\,\text{GHz}}>3\,$mJy/beam & $40,621$ & $-0.70$ & $-0.65$\\
         WISE counterpart, $\Psi< 0.5''$ and $S_{3\,\text{GHz}}>3\,$mJy/beam & $22,235$ & $-0.63$ & $-0.27$\\
         No WISE counterpart, $\Psi< 0.5''$ and $S_{3\,\text{GHz}}>3\,$mJy/beam & $5,769$ & $-0.81$ & $-0.59$\\
    \end{tabular}
    \label{tab:specindex}
\end{table*}

Although our spectral index distribution is cleaner at $S_{3\,\text{GHz}} > 3\,$mJy/beam, we note that the skew to flatter spectral indices when including components below the brightness threshold (see the upper panel of Figure \ref{fig:specindex}) may be a real effect.
If this observation were to hold up with higher quality data, it could support the observations of e.g., \citet{Prandoni2006}, \citet{Owen2008}, and \citet{Whittam2013} of a flattening of spectral indices at flux densities of $\sim 1\,$mJy, likely due to an increase in radio core dominance at low radio luminosities \citep{Whittam2017}.
The future availability of cleaner VLASS \textit{Single Epoch} images will allow reliable measurements of flux densities at this level, and the stacked three-epoch VLASS images will allow testing of this hypothesis using components with $\lesssim 1\,$mJy.

Notably, even when applying a cut on minimum flux density, the spectral index distributions for VLASS components have a skew toward flatter values.
A potential explanation for this is the high resolution of VLASS results in radio lobes being resolved as separate components to their cores.
To investigate the impact of resolved radio cores on the VLASS/FIRST spectral index distribution, we split our sample into likely compact and extended components with a peak brightness of $S_{3\,\text{GHz}} > 3\,$mJy/beam.
We define compact as having $\Psi < 0''.5$ and extended as having $2''.5 < \Psi < 10''$ (and thus larger than a typical VLASS beam). 
Here we find that the spectral index distribution for compact components peaks at $\alpha = -0.66$ with a median of $\alpha = -0.38$, compared to $-0.70$ and $-0.65$ respectively for extended components (see Figure \ref{fig:specindex}).

Furthermore, we can improve the identification of radio cores by requiring a compact component is spatially coincident with an infrared or optical source.
To this end we cross match our sample with the AllWISE catalog \citep{Cutri2013}, based on observations from the Wide-field Infrared Survey Explorer telescope \citep[WISE,][]{Wright2010}.
If a component has an AllWISE detection in the W1 band ($\lambda \sim 3.4\mu$m) within $2.5''$ we associate the radio component and the infrared source with one another. 
Conversely, we consider a radio component to not have an AllWISE match if there is no W1-band detection within 10".
For compact components with an AllWISE match, the spectral index distribution peaks at $\alpha = -0.63$ and has a median of $\alpha = -0.27$.
Where a compact component is not associated with an AllWISE match the spectral index distribution peaks at $\alpha = -0.81$, with a median of $\alpha = -0.59$.
Although this shows that radio cores contribute to the flatter spectrum bias of the VLASS/FIRST spectral index distribution, no account is taken in determining the spectral indices of potential for radio variability that may be present in a few percent of radio sources \citep[e.g.][]{Carilli2003, Mooley2016, Nyland2020}.
The availability of second-epoch VLASS observations will enable the impact of radio variability on this distribution to be quantified.
The peak and median values of all the spectral index distributions analysed here, including those with and without an AllWISE counterpart, are listed in Table \ref{tab:specindex}.



\subsection{Spectral Curvature Between 150 MHz and 3 GHz}
\label{ssec:speccurve}
The availability of the first data release of LoTSS \citep[DR1,][]{Shimwell2019} allows for comparisons of our catalog with deep ($\text{rms}\sim 70\,\mu$Jy/beam), high resolution ($\sim 6''$), low-frequency ($\nu \sim 150\,$MHz) observations over $\sim 400\,\text{deg}^{2}$.
LoTSS DR1 lies within the FIRST footprint, allowing us to match our cross-matched VLASS/FIRST components with LoTSS, again using a $2.5''$ search radius, providing flux density measurements in three radio bands for these components.
We find $\sim 17,000$ VLASS components with observations from both FIRST and LoTSS, and a FIRST deconvolved size smaller than $10''$, $5,497$ of which have $S_{3\,\text{GHz}} > 3\,$mJy/beam.

\begin{figure}
    \centering
    \includegraphics[width=1\columnwidth]{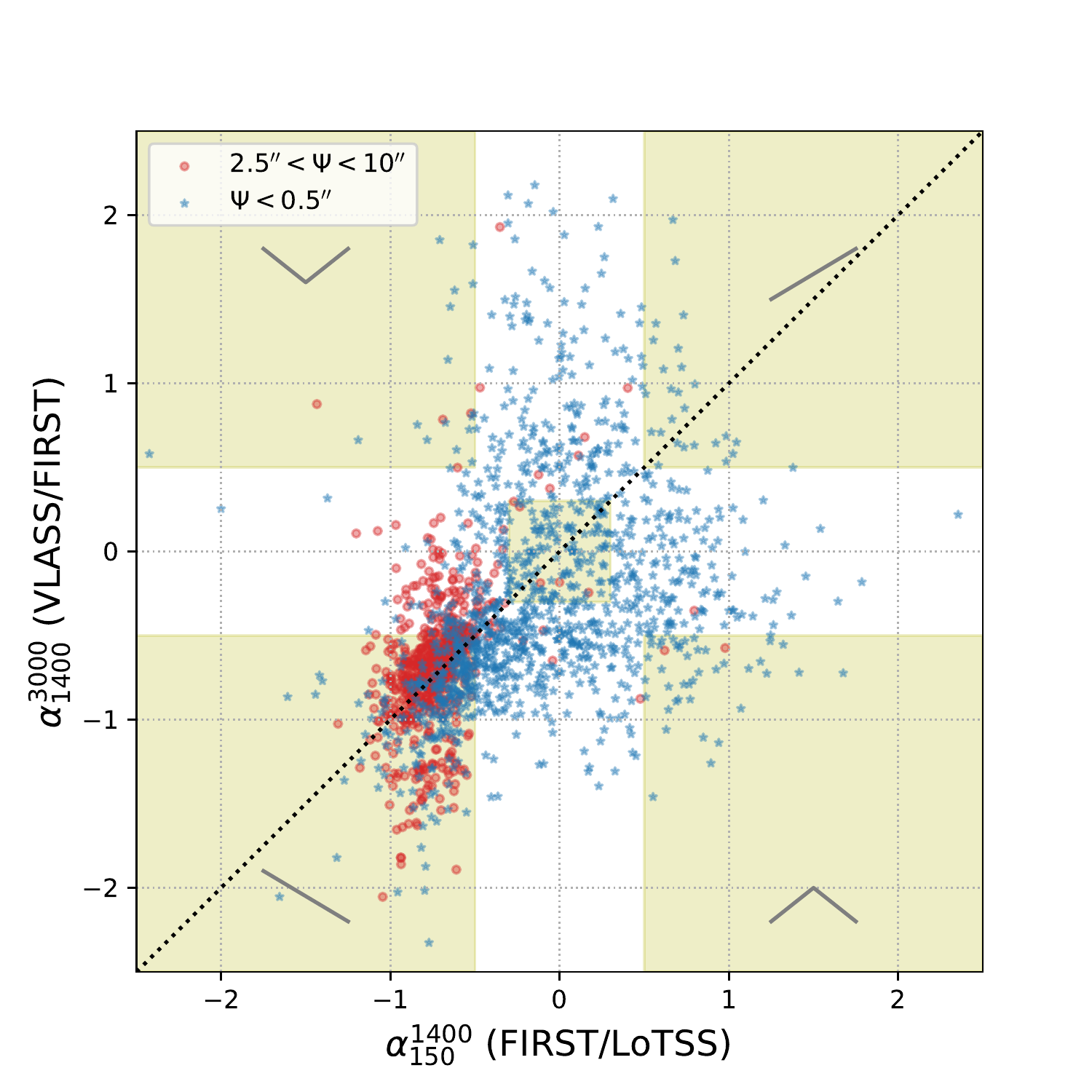}
    \caption{Radio color-color plot for our VLASS components matched with components from both FIRST and LoTSS, where the determined spectral indices are based on the component's total flux density from each survey.
    Blue stars represent compact components ($\Psi < 0.5''$) while extended components ($2.5'' \leq \Psi \leq 10''$) are shown as red circles.
    The x-axis gives the spectral index determined from FIRST ($1,400\,$MHz) and LoTSS ($150\,$MHz), and the y-axis shows the spectral index based on VLASS ($3,000\,$MHz) and FIRST measurements ($1,400\,$MHz).
    The black dotted line on both panels shows $\alpha_{150}^{1400} = \alpha_{1400}^{3000}$.
    The shaded yellow regions show the definitions used in this work for concave-spectrum (upper-left), normal-spectrum (lower-left), inverted-spectrum (upper-right), and GPS (lower-right) sources.
    The grey line in each of these shaded corners shows the representative spectral shape (note that spectral curvature will increase away from the black dotted line).
    The shaded yellow region at the centre defines sources we consider as flat-spectrum.}
    \label{fig:radcolorcolor}
\end{figure}

As in Section \ref{ssec:spidx}, we split our sample of radio components into \textit{compact} (deconvolved size, $\Psi, < 0.5''$, $n=1,205$) and \textit{extended} ($\Psi > 2.5''$, $n=645$) using the deconvolved major axis size of the VLASS component.
As in Section \ref{ssec:spidx} the spectral indices are determined based on the cataloged total flux densities of the components.
Comparing the low- ($150 \lesssim \nu \lesssim 1,400\,$MHz) and high-frequency ($1,400 \lesssim \nu \lesssim 3,000\,$MHz) spectral indices of these components demonstrates their spectral curvature and is shown on a radio `color-color' plot in Figure \ref{fig:radcolorcolor}.
Most components possess normal radio spectra, with the highest density of both compact and extended components at $-1 \lesssim \alpha_{1400}^{3000} \approx \alpha_{150}^{1400} \lesssim -0.5$.
A population of the extended components shows clear spectral curvature, a feature often being associated with older radio lobes \citep{Kharb2008, Harwood2017}.
The radio colors of compact components show substantial scatter, with radio colors displaying inverted spectra (positive spectral index at low- and high-frequency),
flat spectra (spectral index $\sim 0$ at low- and high-frequency), Gigahertz Peaked Spectra \citep[GPS; positive low-frequency and negative high-frequency spectral indices,][]{ODea1998, ODea2020}, and concave spectra (negative low- and positive high-frequency spectral indices).
These distinct populations are highlighted in Figure \ref{fig:radcolorcolor}.
Some of the scatter in the radio color-color parameter space will be driven by radio variability, the impact of which will be quantifiable with the forthcoming availability of VLASS epoch 2 observations.


\subsection{Infrared Colors of Compact Radio Sources With Differing Spectral Shapes}

Considering point-like radio components as individual compact sources can have tangible scientific benefit, allowing us to cross match with multiwavelength observations, and provide insights into the properties of their host galaxies.
To this end we use the cross-match between VLASS and AllWISE described in Section \ref{ssec:spidx} to obtain infrared magnitudes for the likely hosts of these compact sources.
From this cross match we then select five samples of compact sources exhibiting different radio spectra: 
\begin{itemize}
    \item GPS sources with $\alpha_{150}^{1400}>+0.5$ and $\alpha_{1400}^{3000}<-0.5$: $41$ objects, $22$ with AllWISE matches,
    
    \item inverted-spectrum sources having $\alpha_{150}^{1400} > +0.5$ and $\alpha_{1400}^{3000} > +0.5$: $25$ objects, $18$ with AllWISE matches,
    
    \item flat-spectrum sources with $-0.3 <  \alpha_{150}^{1400} < +0.3$ and $-0.3 < \alpha_{1400}^{3000} < +0.3$: $129$ objects, $94$ with AllWISE matches,
    
    \item concave-spectrum sources with $\alpha_{150}^{1400} < -0.5$ and $\alpha_{1400}^{3000} > +0.5$: $17$ objects, $14$ with AllWISE matches,
    
    \item and, for reference, normal-spectrum sources having $\alpha_{150}^{1400} < -0.5$ and $\alpha_{1400}^{3000} < -0.5$: $246$ objects, $147$ with AllWISE matches.
\end{itemize}

\begin{figure*}
    \centering
    \includegraphics[width=2\columnwidth]{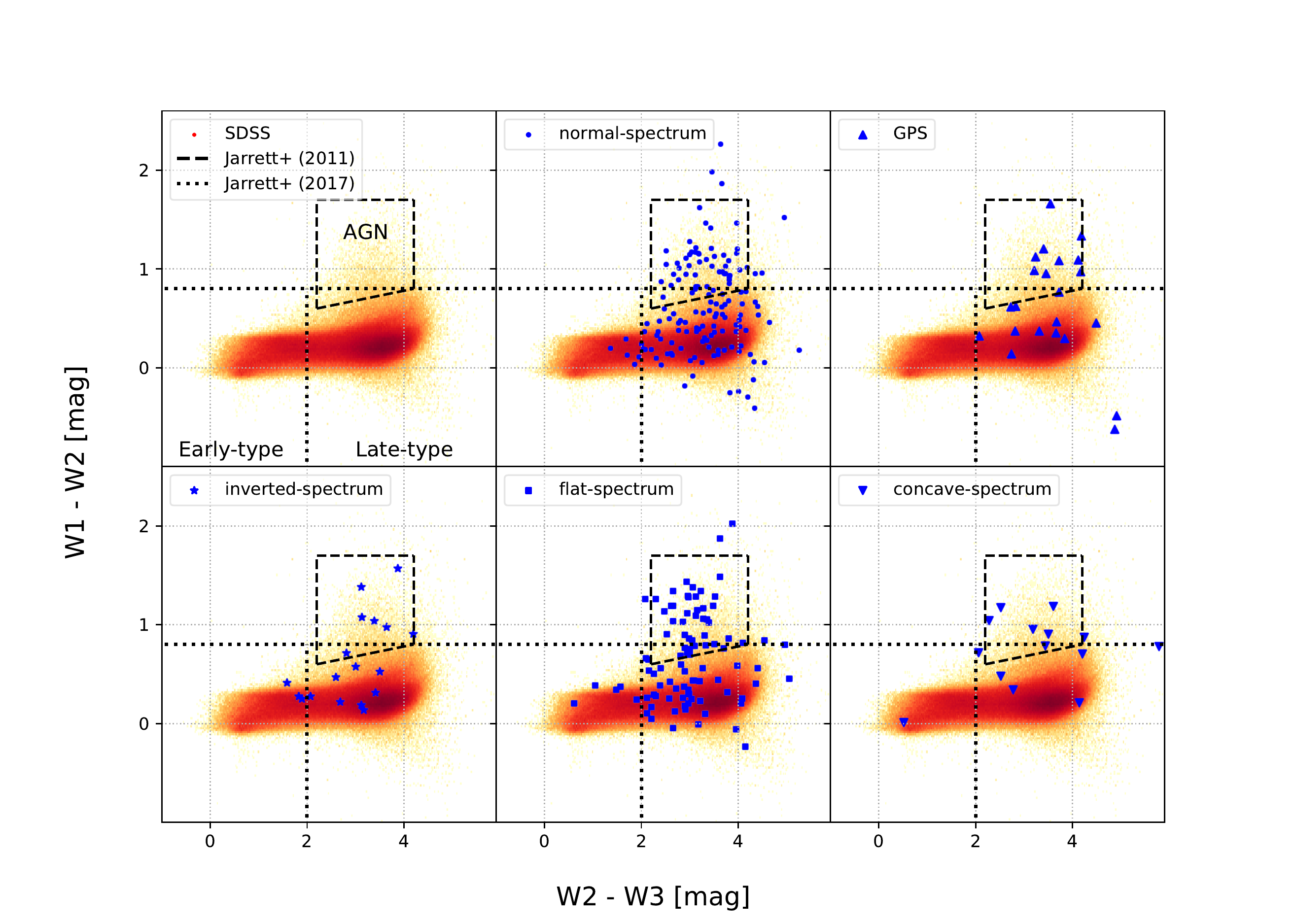}
    \caption{WISE color-color diagrams for compact radio sources with different spectral properties.
    We show five types of radio source here: normal-spectrum sources (blue circles, upper middle panel), GPS sources (blue triangles, upper right), inverted-spectrum sources (blue stars, lower left), flat-spectrum sources (blue square, lower middle), and concave-spectrum sources (inverted blue triangles, lower right).
    The WISE color-color distribution of galaxies in SDSS is shown in shown red for reference (and for clarity on its own in the upper-left panel).
    The $x$-axis shows the $\text{W}1 - \text{W}2$ color, and $\text{W}2 - \text{W}3$ is shown on the $y$-axis.
    The black-dashed line shows the quasar/AGN region defined by \citet{Jarrett2011}, and the black dashed line marks out the regions of color-color space occupied by early- (left) and late-type galaxies \citep[right,][]{Jarrett2017}.
    }
    \label{fig:wisecolors}
\end{figure*}

In Figure \ref{fig:wisecolors} we show the WISE colors ($\text{W}1\ [3.4\,\mu\text{m}] - \text{W}2\ [4.3\,\mu\text{m}]$ and $\text{W}2 - \text{W}3\ [12\,\mu\text{m}]$) of the samples defined above.
We then classify the host galaxies as either early- or late-type using the criteria of \citet{Jarrett2017}, or as AGN based on \citet{Jarrett2011}.

Infrared colors associated with early-type galaxies are relatively uncommon amongst our sample \textemdash\ potentially the result of only including unresolved radio sources \textemdash\ accounting for $3.4_{-0.9}^{+2.2}\,\%$ of normal-spectrum, $5.3_{-1.5}^{+3.3}\,\%$ of flat-spectrum,
and $7.1_{-2.3}^{+13.2}\,\%$ of concave-spectrum sources.
None of the 26 GPS sources in our sample have WISE colors expected of early-type galaxies.
For inverted-spectrum sources however, $16.7_{-5.4}^{+12.1}\,\%$ have WISE colors associated with early-type hosts.
Unfortunately, the small size of our sample (17 objects) prevents any statistically significant conclusions being drawn here, but we note that where inverted-spectrum sources are associated with early-type colors, these colors are within half a magnitude of the \citet{Jarrett2017} early-/late-type segregation of the WISE color-color space.

More generally, the WISE colors of our sample of compact radio sources are split between being consistent with late-type and AGN hosts.
Late-type hosts account for $43 - 57\,\%$ of compact radio sources (depending on spectral shape), whilst $35 - 45\,\%$ of sources have AGN-like WISE colors.
The fractions of normal-spectrum, GPS, inverted-, flat- and concave-spectrum sources having WISE colors associated with AGN, late- and early-type galaxies are explicitly given in Table \ref{tab:wcolors}.
For comparison with the general galaxy population we also show the WISE colors of MPA/JHU sample from the seventh data release of the Sloan Digital Sky Survey \citep[SDSS,][]{York2000, Kauffmann2003, Abazajian2009} in Figure \ref{fig:wisecolors}, and the fractions of each color classification (AGN, late- or early-type) are given in Table \ref{tab:wcolors}. 
It is worth noting however, that while the host of a radio source may have WISE colours associated with star-forming disk galaxies as defined by \citet{Jarrett2017}, that in and of itself does not mean that star-formation is responsible for the radio emission as opposed to an AGN.


\begin{table}
    \centering
    \caption{Percentage and binomial errors of fractions of compact sources with IR counterpart classifications as defined by \citet{Jarrett2011, Jarrett2017} based on their WISE colors (see also Figure \ref{fig:wisecolors}).}
    \begin{tabular}{l|c c c}
         \textbf{Sample} & \textbf{AGN} & \textbf{late-type} & \textbf{early-type}\\
          & \textbf{[$\%$]} & \textbf{[$\%$]} & \textbf{[$\%$]}\\
         \hline 
         SDSS & $0.9_{-0.01}^{+0.01}$ & $75.4\pm0.06$ & $23.6\pm0.05$\\
         normal-spectrum & $35.4_{-3.7}^{+4.1}$ & $57.1_{-4.2}^{+4.0}$ & $3.4_{-0.9}^{+2.2}$\\
         GPS & $45.5_{-9.8}^{+10.6}$ & $54.5_{-10.6}^{+9.8}$ & $0.0^{+7.7}$\\
         inverted-spectrum & $38.9_{-9.8}^{+12.1}$ & $44.4_{-10.6}^{+11.7}$ & $16.7_{-5.4}^{+12.1}$\\
         flat-spectrum & $41.5_{-4.9}^{+5.2}$ & $48.9\pm5.1$ & $5.3_{-1.5}^{+3.3}$\\
         concave-spectrum & $42.9_{-11.5}^{+13.3}$ & $42.9_{-11.5}^{+13.3}$ & $7.1_{-2.3}^{+13.2}$\\
    \end{tabular}
    \label{tab:wcolors}
\end{table}

Despite the clear advantages of matching the VLASS, FIRST and LoTSS catalogs, the observations were not made simultaneously.
Thus, there will exist contamination from sources exhibiting radio variability.
Tackling this is beyond the scope of the work presented here which acts as a small-scale proof of concept. 
However, the future advent of multi-epoch observations from VLASS will be invaluable in identifying such detections.
Additionally as more data is released by LoTSS the footprint over which such radio color-color data is available will increase beyond the $\sim 400\,\text{deg}^{2}$ of the first LoTSS data release utilised here.
Eventually such a VLASS/FIRST/LoTSS color-color map will be limited only by the $\sim 10,000\,\text{deg}^{2}$ of the FIRST footprint.



\section{Euclidean Normalised Source Counts}
\label{sec:vlassdnds}

For the first time wide-field coverage of the sky at $3\,$GHz down to mJy sensitivities is made available through this data.
Thus this presents a prime opportunity to quantify the Euclidean normalised (i.e. normalised to the assumption of a geometrically flat and steady state Universe) $3\,$GHz source counts, $dN/dS$, at this depth.
In order to do this we must account for several complicating factors.

Firstly, in Figure \ref{fig:aitoffrms} we demonstrate that the depth of the \textit{Quick Look} imaging is distinctly non-homogeneous.
This will impact upon the effective survey footprint over which faint sources can be detected.
That is to say, that whilst a $1\,$mJy component may be readily detected in a relatively deep part of the survey, in one of the more shallow regions (e.g. the Galactic plane which is shown in Figure \ref{fig:aitoffrms} to be particularly noisy) that component may not be detected.
The dominant source of noise variance in the \textit{Quick Look} images is due to observing tile edge effects \textemdash\ the distinctly higher noise at the edge of a VLASS tile.
To estimate the impact of this we compare the peak brightness distributions of components more than $0.2^{o}$ from a tile edge (and therefore not impacted by tile edge effects) with the full VLASS component peak brightness distribution.
Bin-wise dividing the not-edge peak brightness distribution by the full sample peak brightness distribution provides the average completeness (and thus incompleteness) as a function of peak brightness due to tile edge-effects, $p_{\text{edges}}(S)$.
Additionally, the average noise in a particular image can be impacted by the presence of bright radio components.
By taking $5\times$ the mean image rms as the limiting flux density for a component to be detected in an image, we can estimate the sensitivity distribution of the \textit{Quick Look} images.
We can combine the distribution of image rms with the completeness due to tile edge effects to give a probability distribution, $P(S)$, of a component of flux density $S$ being detected across the survey footprint, i.e., $P(S) = p_{\text{edges}}(S) \times p_{\text{imdepth}}(S)$, where $p_{\text{edges}}(S)$ is the contribution from tile edge effects, and $p_{\text{imdepth}}(S)$ is the contribution from varying image effective depth.
At $S_{\text{peak}}=3\,$mJy/beam the sample completeness is $99.8\,\%$, whereas at the $1\,$mJy/beam level the completeness of the catalog is $67\,\%$.
The \textit{effective} survey footprint as a function of component brightness is then estimated by $P(S)\times A_{\text{VLASS}}$, where $A_{\text{VLASS}}$ is the total observed footprint of the survey.

\begin{figure}
    \centering
    \subfigure{\includegraphics[trim=0 2mm 0 0, clip, width=\columnwidth]{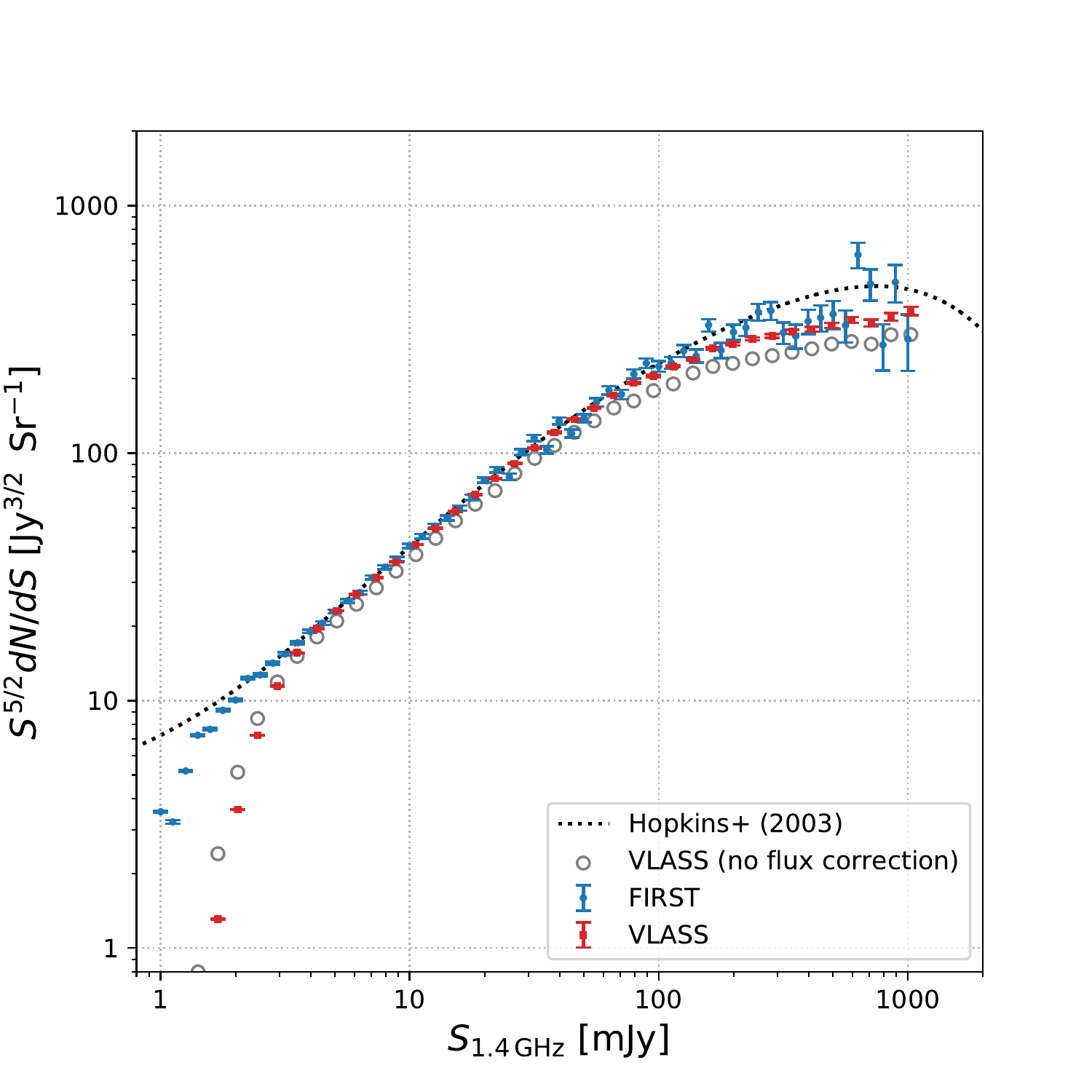}}
    \hspace{3mm}
    \subfigure{\includegraphics[trim=0 2mm 0 18mm, clip, width=\columnwidth]{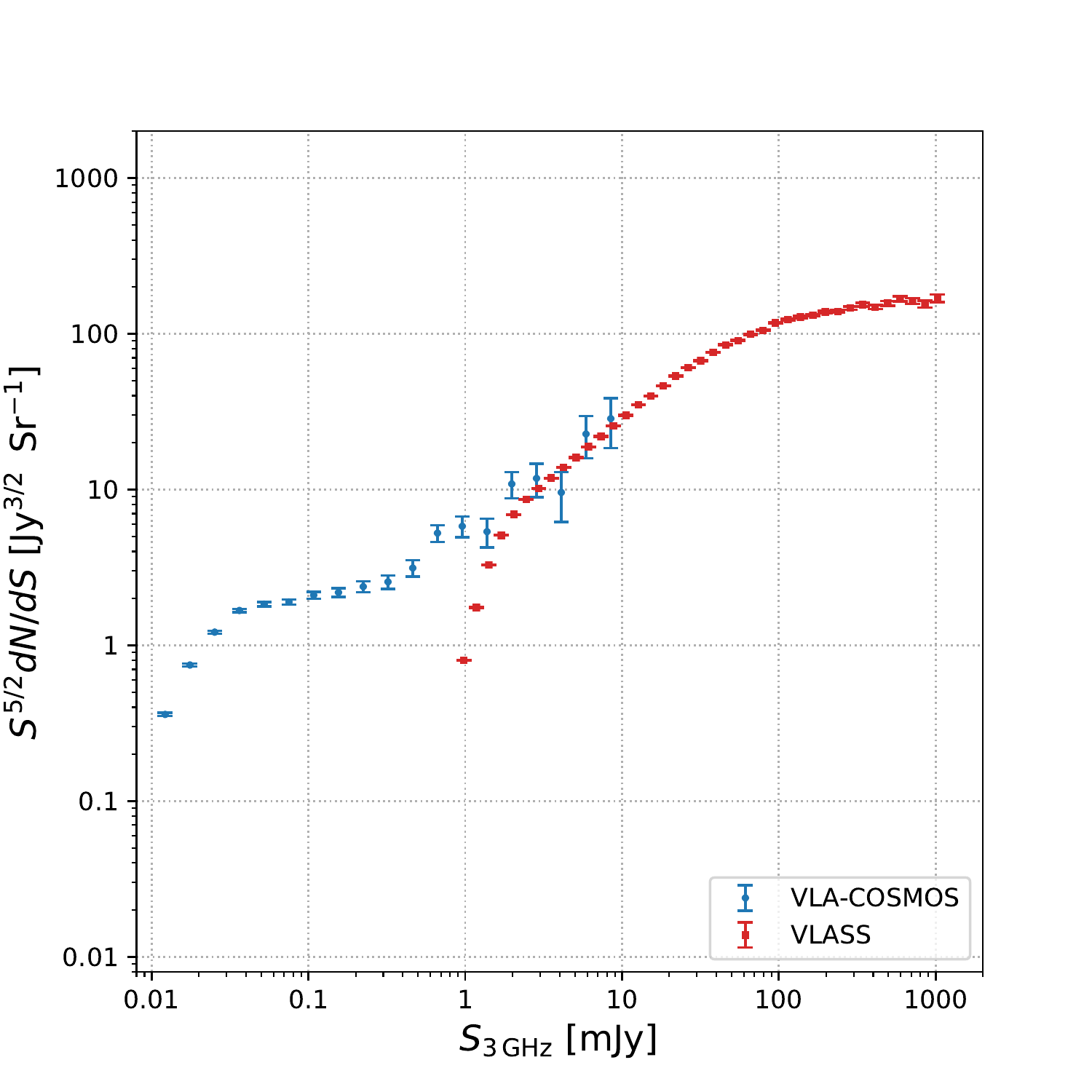}}
    \caption{
    The Euclidean normalised differential source counts, $dN/dS$, for VLASS.
    The upper panel shows VLASS converted to $1.4\,$GHz for comparison with FIRST (blue circles).
    The red squares show the VLASS source counts after scaling the cataloged flux density by $15\,\%$ as derived in Section \ref{sssec:fluxcal}.
    The grey empty circles show the VLASS $dN/dS$ without correcting for the VLASS flux underestimate. 
    For reference the black dotted line shows the sixth-order polynomial fit derived by \citet{Hopkins2003}.
    In the lower panel we show the VLASS $dN/dS$ at $3\,$GHz (red squares) alongside the VLA-COSMOS source counts (blue circles).}
    \label{fig:logNlogS}
\end{figure}

The second complication in deriving the VLASS \textit{Quick Look} $dN/dS$ is due to the fact that individually detected radio components are often resolved parts of larger structures that make up a distinct source. 
To counter this, and for consistency with previous works, we adopt the approach that most radio components within $50''$ are related \citep{Oosterbaan1978, Windhorst1985, White1997}.
Hence, for all components within $50''$ of one another their integrated flux densities are summed and they are treated as a single source.

The VLASS \textit{Quick Look} $dN/dS$, corrected for $P(S)$,  is shown in Figure \ref{fig:logNlogS} alongside VLA-COSMOS \textemdash\ the previously largest available $3\,$GHz dataset.
Whilst VLA COSMOS probes deeper than VLASS, there is overlap in the flux density range covered by these surveys between \text{$1 \lesssim S \lesssim 10\,$mJy}, and above $\sim3\,$ mJy the VLASS $dN/dS$ is shown to be consistent with the $dN/dS$ from VLA-COSMOS.
To compare with a radio survey covering a similar flux density range we additionally compare the VLASS $dN/dS$ to the FIRST $dN/dS$ (also shown in Figure \ref{fig:logNlogS}) by extrapolating the VLASS flux measurements to $1.4\,$GHz assuming a spectral index of $\alpha=-0.71$.
An additional benefit to comparing to the FIRST $dN/dS$ is that the radio source counts at $1.4\,$GHz have been well quantified and modelled by numerous observations from Jy level down to sub-mJy depths \citep[e.g.,][]{Windhorst1985, White1997, Hopkins2003}. 
Here we find that at $S_{1.4\,\text{GHz}}\gtrsim 5\,$mJy the  VLASS \textit{Quick Look} $dN/dS$ is consistent with both the FIRST differential source counts and the 6th-order polynomial fit to the $1.4\,$GHz $dN/dS$ obtained by \citet{Hopkins2003}.
Assuming $\alpha = -0.71$, $5\,$mJy at $1.4\,$GHz is equivalent to $2.9\,$mJy at $3\,$GHz.
The agreement with the $1.4\,$GHz source counts is also indicative that the VLASS flux calibration derived in Section \ref{sssec:fluxcal} is accurate, and we show the $1.4\,$GHz $dN/dS$ for VLASS using uncorrected flux measurements on this same plot for reference.
The comparison with the $dN/dS$ at both $3$ and $1.4\,$GHz as established by previous works is consistent with the VLASS \textit{Quick Look} catalog being incomplete at flux densities below $S_{3\,\text{GHz}} \sim 3\,$mJy.

\section{The Sizes of Radio Sources}
\label{sec:angsizes}

\subsection{Angular Size distribution of VLASS Components}

\begin{figure}
    \centering
    \subfigure{\includegraphics[trim=0 0mm 0 0, clip, width=\columnwidth]{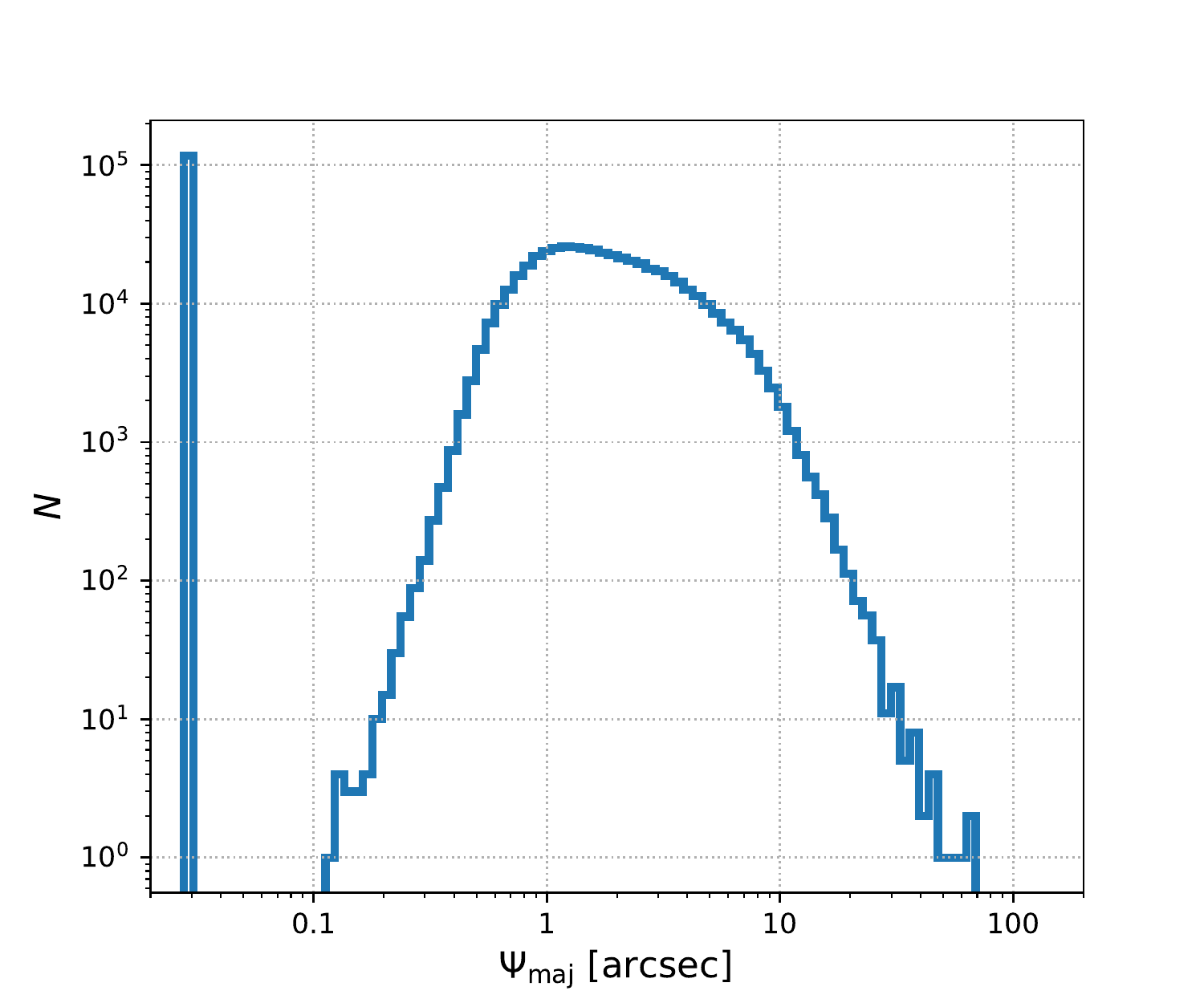}}
    \subfigure{\includegraphics[trim=0 0mm 0 15mm, clip, width=\columnwidth]{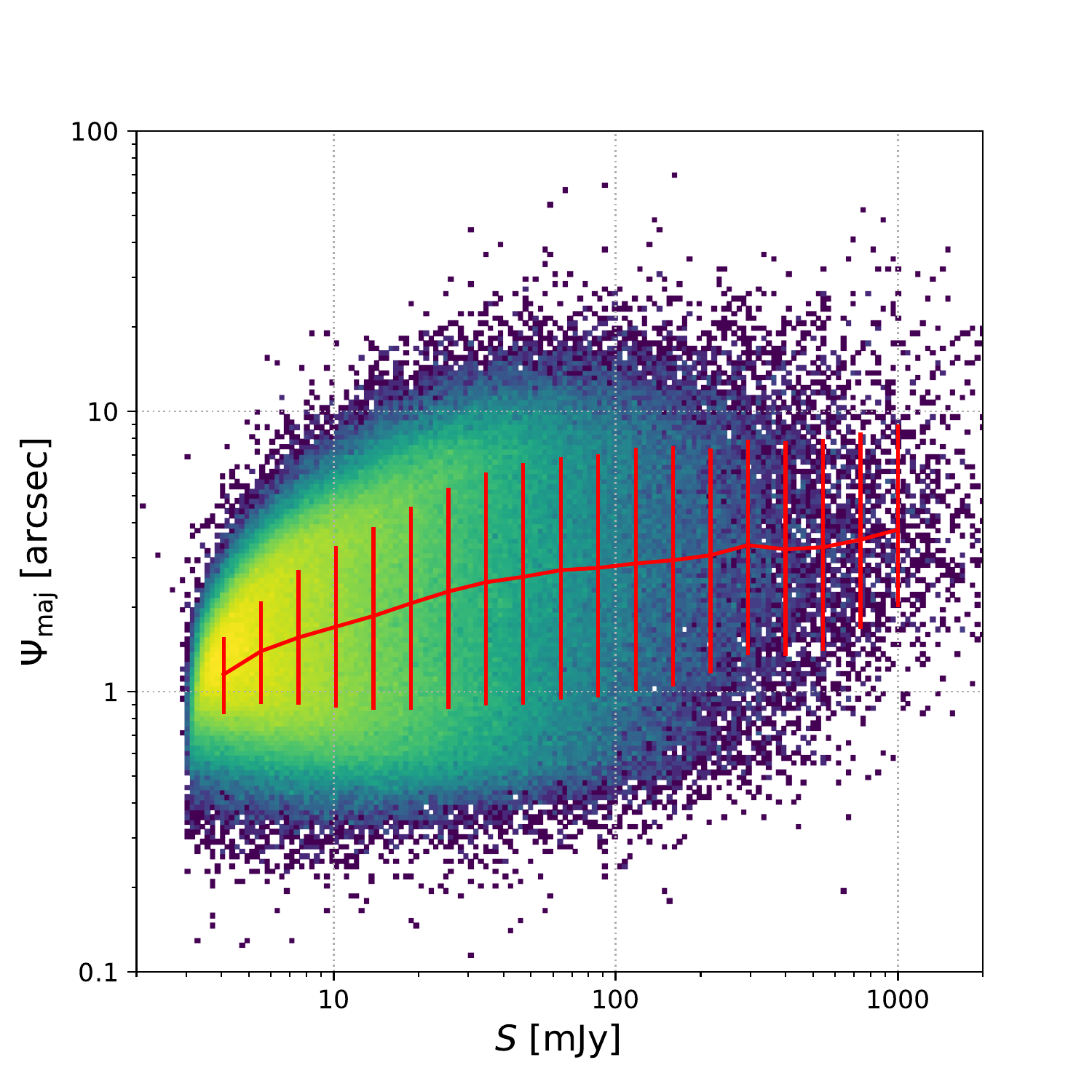}}
    \caption{The angular size distributions of VLASS components with $S_{\text{peak}}>3\,$mJy/beam.
    The upper panel shows a histogram of the deconvolved major axis size, $\Psi_{\text{maj}}$. 
    The number of components with a size listed as zero in the catalog (i.e., smaller than can be measured by VLASS) are shown in the bin at $0''.03$.
    The lower panel shows the distribution of the deconvolved major axis for VLASS components with a non-zero cataloged size versus component total flux density.
    The red line shows the median deconvolved angular size of the component for a particular flux bin.
    Error bars are defined by the $16$th and $84$th percentiles of the size distribution in each flux bin, and broadly shows a weak ($r = 0.3$) correlation with flux with $\Psi \propto S^{0.3}$ below 50 mJy.}
    \label{fig:sizevflux}
\end{figure}

With a median uncertainty of $\sim0''.1$, the distribution of deconvolved component sizes, $\Psi$, in VLASS probes down to a scale of $\sim 0''.3 - 0.''4$.
Below this, components are listed as having a zero size as observed in VLASS, and higher resolution observations are required to measure their sizes.
The distribution of the full-width half maximum of deconvolved major axis, $\Psi_{\text{maj}}$, of VLASS components with $S_{\text{peak}} > 3\,$mJy/beam is shown in Figure \ref{fig:sizevflux}.
Here we denote the population of components where $\Psi_{\text{maj}}$ is listed as zero in the bin at $\Psi_{\text{maj}} = 0''.03$.

For those components with a measured, rather than zero, deconvolved angular size we can exploit the availability of this large high-resolution data set to investigate the relation between radio source size and total flux density \citep{Windhorst1990} at $\nu \sim 3\,$GHz.
In the lower panel of Figure \ref{fig:sizevflux} we show the distribution of non-zero deconvolved VLASS component sizes as a function of total component flux.
Our sample shows substantial scatter,
but with a trend for increasing size with flux density.
For $S_{\text{total}} \lesssim 50\,$mJy the median component size follows a trend of $\Psi(S) \propto S^{0.3\pm0.01}$, consistent with $\Psi = 2''S_{1.4\,\text{GHz}}^{0.3}$ reported by \citet{Windhorst1990}.
For VLASS components brighter $S\gtrsim 50\,$mJy the slope of this trend flattens somewhat to $\sim0.2$.
However, this may be attributable to the VLASS survey design. 
VLASS is run in the VLA's B and BnA configurations\footnote{see \url{https://science.nrao.edu/facilities/vla/docs/manuals/propvla/array_configs} for more detail on VLA observing configurations.}.
One of the impacts of this observing mode is that sensitivity to large angular scale structure is reduced by a lack of short baselines used in the array configuration.
In the case of VLASS this results in larger ($\Psi \gtrsim 30''$) objects being resolved into multiple components or lost altogether \citep{Lacy2020}. 
As a consequence, VLASS will only sample components with small angular sizes at higher flux densities.
A further effect is that the relationship between linear size and flux density is dependent on a source not being split into multiple components. 
Consequently, components detected in a high angular resolution such as VLASS are only suitable for testing this relation at small angular scales, and testing at large angular scales necessitates a sample of verified multi-component sources.


\subsection{The VLASS Two-Point Correlation Function}
\label{sec:twopoint}

The high resolution of VLASS lends itself to exploring small scale clustering of radio detections.
To this end we determine the angular two-point correlation function for VLASS using the \citet{Landy1993} estimator:
\begin{equation}
    \xi_{\text{LS}} = \frac{DD-2DR+RR}{RR},
\end{equation}
where $DD$ is the real data autocorrelation, $RR$ is the autocorrelation of random data, and $DR$ is the cross correlation of real and random data.
Given the angular resolution of VLASS is about twice that of FIRST one might expect the two-point correlation functions of these surveys to diverge at angular scales below the FIRST beam size.
To that end we additionally determine the FIRST two-point correlation function for comparison.

\begin{figure}
    \centering
    \includegraphics[width=\columnwidth]{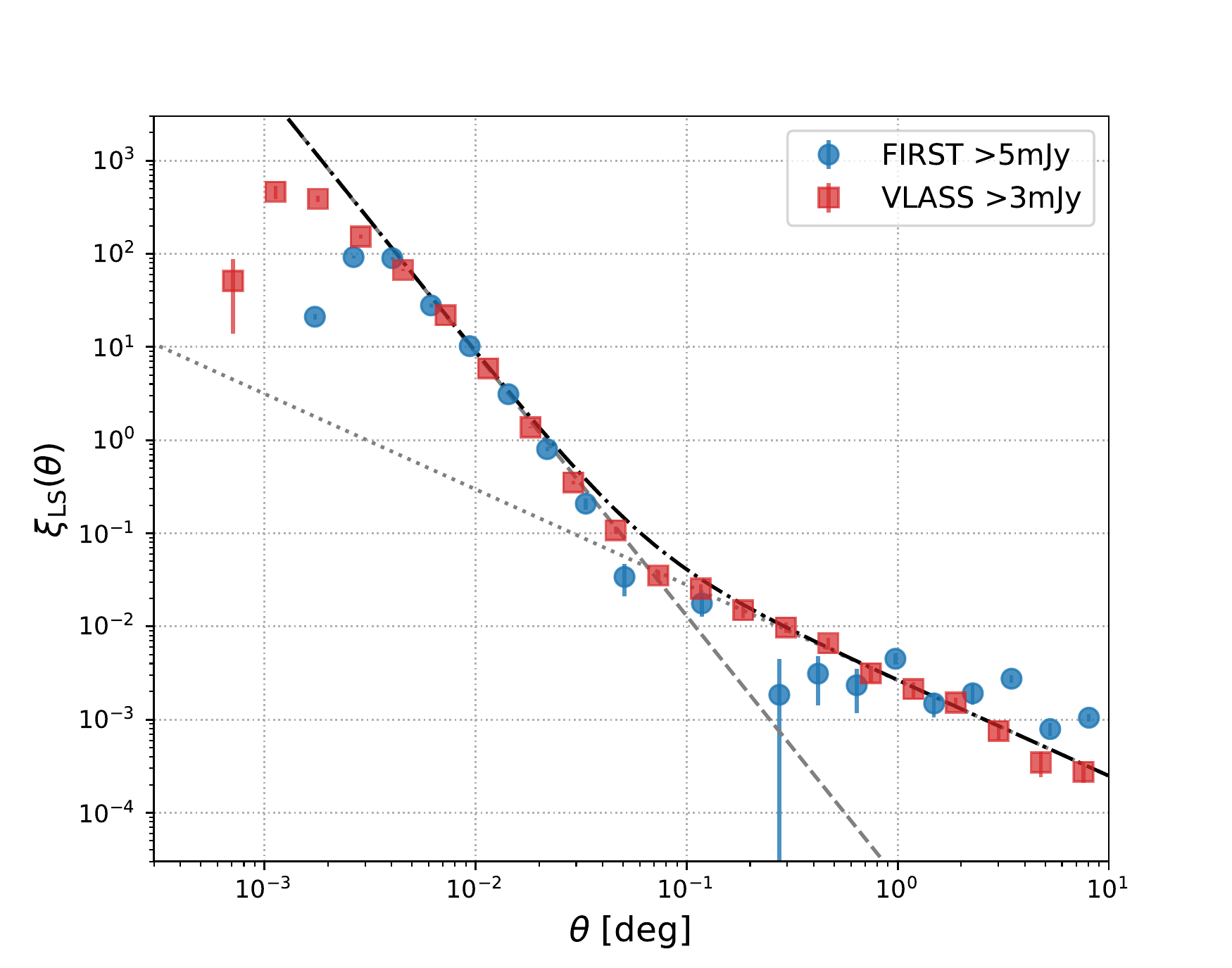}
    \caption{Two-point correlation for VLASS components with $S_{\text{Total}} > 3\,$mJy (red squares).
    For comparison the two-point correlation function for FIRST is also shown (blue circles).
    The VLASS two-point correlation shows excess power at $\theta \lesssim 0.002^{o} (\sim 7'')$ relative to FIRST, as expected given the higher resolution of VLASS.
    The black dashed line shows the sum of the two power laws fit to the VLASS two-point correlation function (see Table \ref{tab:twopointparams}), while the grey dot-dashed and dotted lines show the individual power laws fitted.}
    \label{fig:twopoint}
\end{figure}

In Figure \ref{fig:twopoint} we show both the VLASS and FIRST two-point correlation functions over the range $0.0003^{\circ} (1.1'') < \theta < 10^{\circ}$.
To compensate for incompleteness at low flux densities only VLASS components with $S_{3\,\text{GHz}} > 3\,$mJy are used in this analysis. 
To ensure a fair comparison with FIRST, we limit our two-point correlation for FIRST to components with $S_{1.4\,\text{GHz}}>5\,$mJy.
We fit the VLASS two-point correlation with two power laws, such that:
\begin{equation}
    \xi(\theta) = A\theta^{\alpha} + B\theta^{\beta}
    \label{eq:twopointeq}
\end{equation}
where, $A\theta^{\alpha}$ and $B\theta^{\beta}$ are power law fits at angular scales larger and smaller than $0.05^{\circ}$ respectively.
For the VLASS \textit{Quick Look} data: $A = 2.6\pm0.3\times 10^{-3}$, $B = 1.9\pm0.3\times 10^{-5}$, $\alpha = -1.03\pm0.03$, and $\beta = -2.83\pm0.10$.
These parameters, as well as our determined values for the FIRST two-point correlation, are shown in Table \ref{tab:twopointparams} alongside the $\chi^{2}$ values for the two power law fits.

\begin{table}
    \caption{Power law parameters for Equation \ref{eq:twopointeq} determined from both VLASS and FIRST, and the $\chi^{2}$ value for the fit to the respective data set.}
    \begin{tabular}{l|c c }
         & \textbf{VLASS} & \textbf{FIRST}\\
         \hline 
         $A [\times 10^{-3}]$ & $2.6\pm0.3$ & $2.0\pm0.3$\\
         $\alpha$ & $-1.03\pm0.03$ & $-0.87\pm0.18$\\
         $B [\times 10^{-5}]$ & $1.9\pm0.3$ & $2.3\pm0.4$\\
         $\beta$ & $-2.83\pm0.10$ & $-2.77\pm0.11$\\
         $\chi^{2}$ & $2.82$ & $1.59$\\
    \end{tabular}
    \label{tab:twopointparams}
\end{table}

The VLASS two-point correlation function shows a slope of $-1.03$ at angular scales greater than $0.05^{o}$.
As expected, the gradient of the VLASS two-point correlation function steepens at $\theta \lesssim 0.05^{o}$ to $-2.83$.
This break is the result of radio sources being split into multiple components at smaller angular scales resulting in excess observed clustering \citep{Cress1996, Blake2002}.
Notably, the VLASS two point correlation function shows excess power relative to FIRST at angular scales smaller than $\theta \sim 0.002^{\circ} (\sim 7'')$, the result of VLASS having twice the angular resolution of FIRST \textemdash\ multi-component sources at this scale split by VLASS are usually observed as a single component by FIRST (see also Section \ref{ssec:splitfirst}).
The excess power at small angular scales in the VLASS two-point correlation function demonstrates the potential to probe the angular size distribution of double radio sources down to the order of a few arcseconds with VLASS.

\subsection{Splitting Close Doubles With VLASS}
\label{ssec:splitfirst}

\begin{figure}
    \centering
    \subfigure{\includegraphics[width=0.3\columnwidth]{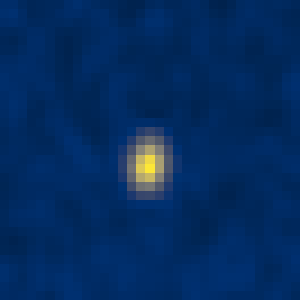}}
    \hspace{1mm}
    \subfigure{\includegraphics[width=0.3\columnwidth]{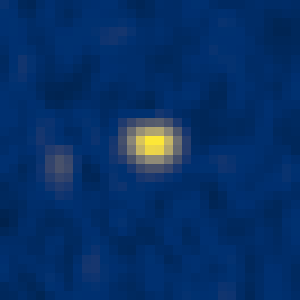}}
    \hspace{1mm}
    \subfigure{\includegraphics[width=0.3\columnwidth]{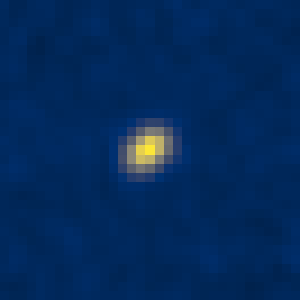}}
    
    \vspace{-3mm}
    
    \subfigure{\includegraphics[width=0.3\columnwidth]{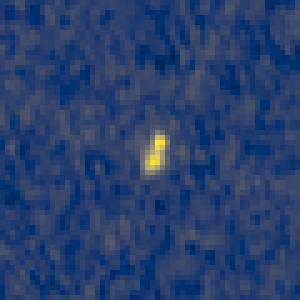}}
    \hspace{1mm}
    \subfigure{\includegraphics[width=0.3\columnwidth]{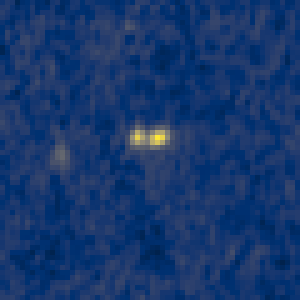}}
    \hspace{1mm}
    \subfigure{\includegraphics[width=0.3\columnwidth]{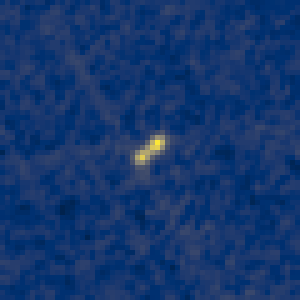}}
    
    
    \subfigure{\includegraphics[width=0.3\columnwidth]{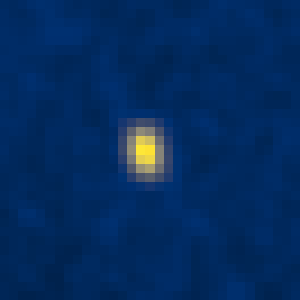}}
    \hspace{1mm}
    \subfigure{\includegraphics[width=0.3\columnwidth]{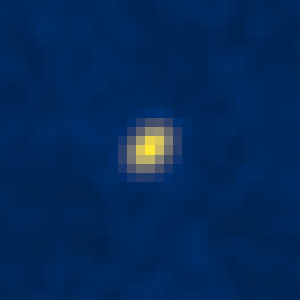}}
    \hspace{1mm}
    \subfigure{\includegraphics[width=0.3\columnwidth]{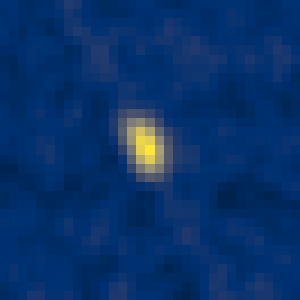}}
    
    \vspace{-3mm}
    
    \subfigure{\includegraphics[width=0.3\columnwidth]{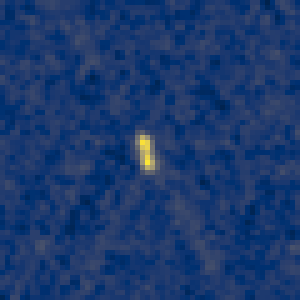}}
    \hspace{1mm}
    \subfigure{\includegraphics[width=0.3\columnwidth]{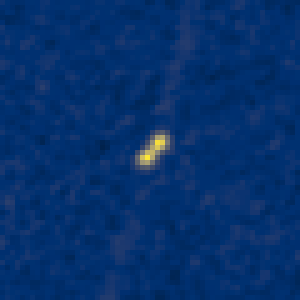}}
    \hspace{1mm}
    \subfigure{\includegraphics[width=0.3\columnwidth]{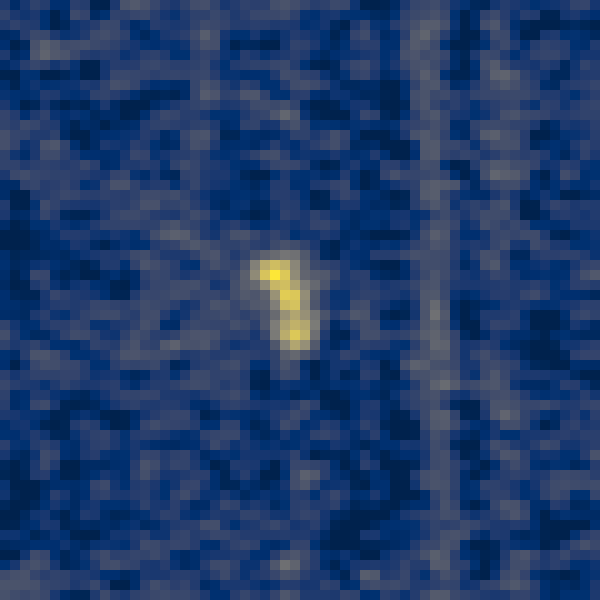}}
    
    \caption{
    Six examples of single-FIRST-component AGN from \citet{Best2012} that are split into multiple components by VLASS.
    The postage stamp images are $1' \times 1'$, and are arranged into two rows and three columns of paired images.
    The top panel in a pair shows the FIRST image and the bottom panel shows the VLASS image.
    Five of the six examples show VLASS doubles, and the lower-right image pair shows an example of single-FIRST-component AGN split into a triple source by VLASS.}
    \label{fig:resbhims}
\end{figure}

Given that VLASS has a similar sensitivity to FIRST but with twice as good an angular resolution, it stands to reason that some fraction of double radio galaxies that are observed as a single component in FIRST will be split into double or triple sources by VLASS.
To test this we take the AGN from the \citet{Best2012} catalog of radio galaxies that are associated with a single FIRST component.
Using this data has the double advantage that a) the radio observations have been associated with a host galaxy already, and b) that the radio emission has been identified as being due to an AGN, rather than star formation and thus is potentially an unresolved double or triple radio source.
To test this we search a radius of $10''$ around each of the $11,556$ single-FIRST-component AGN in the \citet{Best2012} catalog counting the number of matches within our catalog for each \citet{Best2012} source - those with multiple VLASS matches are likely to be sources that have been split into multiple components by VLASS.
In Figure \ref{fig:resbhims} we show six examples of single-FIRST-component AGN that are split into multiple components by VLASS (five doubles and one triple). 

\begin{figure}
    \centering
    \includegraphics[width=\columnwidth]{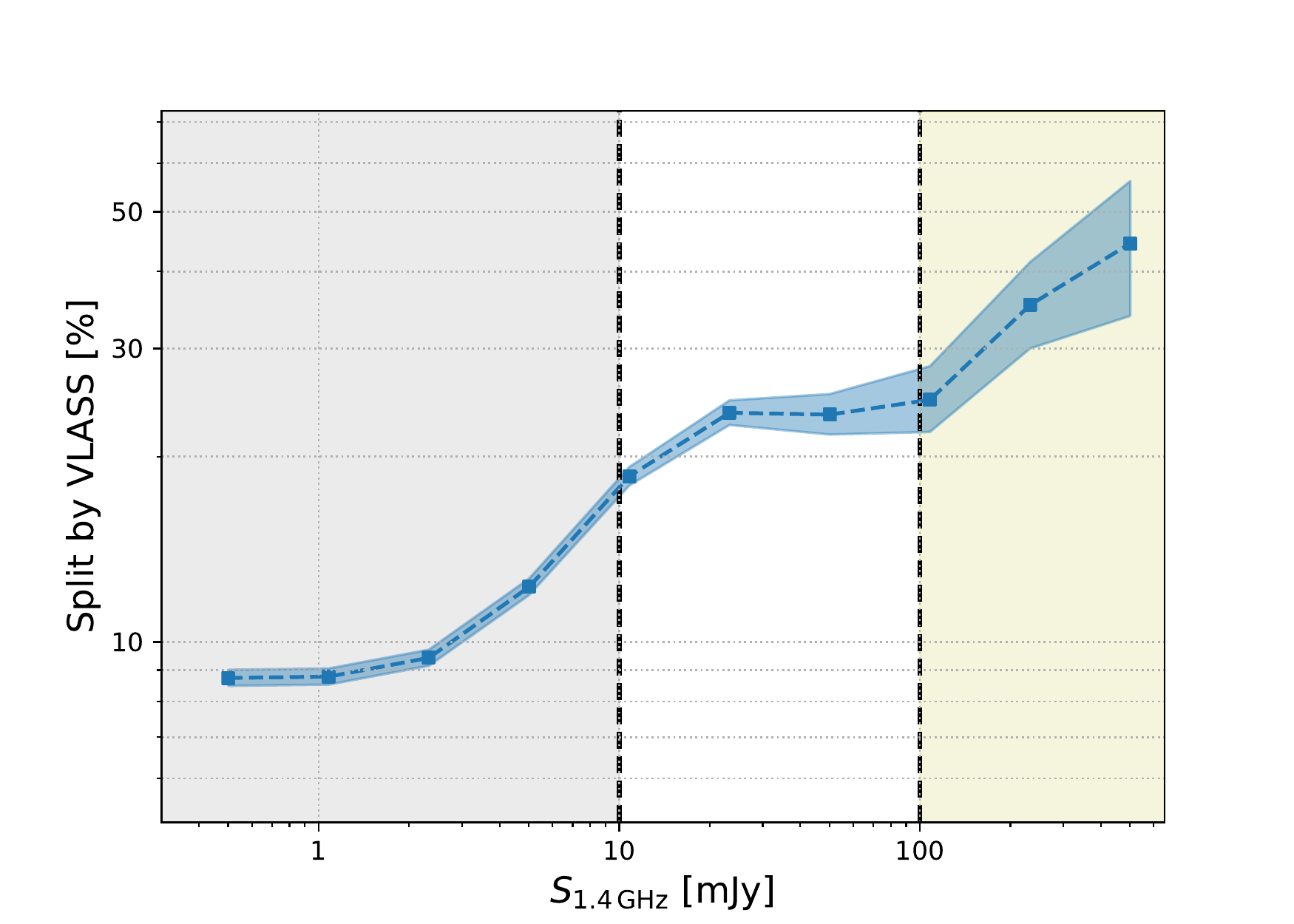}
    \caption{The percentage of AGN in the radio galaxy catalog of \citet{Best2012} associated with a single FIRST component that have been split into multiple components by VLASS plotted as a function of $1.4\,$GHz flux density.
    The grey shaded region to the left of both black dashed lines represents $S_{1.4\,\text{GHz}} < 10\,$mJy, where the combination of spectral index and spreading the emission across multiple components is expected to result in missed VLASS detections.
    The yellow shaded region to the right of both black dashed lines shows $S_{1.4\,\text{GHz}} > 100\,$mJy, where the high flux density of VLASS components may result in sidelobes contaminating the fraction of multi-VLASS-single-FIRST-component sources.}
    \label{fig:resBH12}
\end{figure}

We find that $1,011$ single-FIRST-component AGN from \citet{Best2012} are split into multiple components by VLASS.
Of these, $890$ objects are split into two components, $110$ sources are split into three components, $9$ are split into four components, and two sources are split into five VLASS components.
In Figure \ref{fig:resBH12} we show the fraction of \citet{Best2012} single-FIRST-component AGN split into multiple components by VLASS as function of the total flux density of that source in the FIRST catalog.
At lower flux densities ($1 \lesssim S_{1.4\,\text{GHz}} \lesssim 10\,$mJy) the fraction of resolved sources increases from $8\,\%$ to $\sim 20\,\%$ with increasing flux density.
This trend of increasing resolved fraction is probably due to sensitivity and frequency differences between the surveys. 
For instance, a typical spectrum radio source with $\alpha = -0.7$ and a $1.4\,$GHz flux density of $1\,$mJy would be expected to have a $3\,$GHz flux density of $0.59\,$mJy.
If that source is then split into multiple components the flux density of each component is naturally lower than the sum of the total source flux density \textemdash\ e.g. a $0.6\,$mJy source could be resolved into two components of $0.3\,$mJy each.
At $10\lesssim S_{1.4\,\text{GHz}} \lesssim 100\,$mJy, around $20\,\%$ of single-FIRST-component AGN are split into multiple components by VLASS.
For sources brighter than this, the resolved fraction tends to increase with flux density again. 
This is likely biased by false-positive detections near bright components that were not removed by our quality flagging.
For example sidelobes around sources brighter than $S_{1.4\,\text{GHz}} > 100\,$mJy are themselves bright enough that they may not be flagged by the catalog quality flagging algorithm. 
Such cases will contribute to a fraction of single-FIRST-component AGN that may appear to have multiple components in the VLASS \textit{Quick Look} images and artificially inflate the resolved fraction.
Taking sources with $10 < S_{1.4\,\text{GHz}} < 100\,$mJy $17.7 \pm 0.6\,\%$ of single-FIRST-component AGN are split into multiple components by VLASS.

\begin{figure}
    \centering
    \subfigure{\includegraphics[ width=\columnwidth]{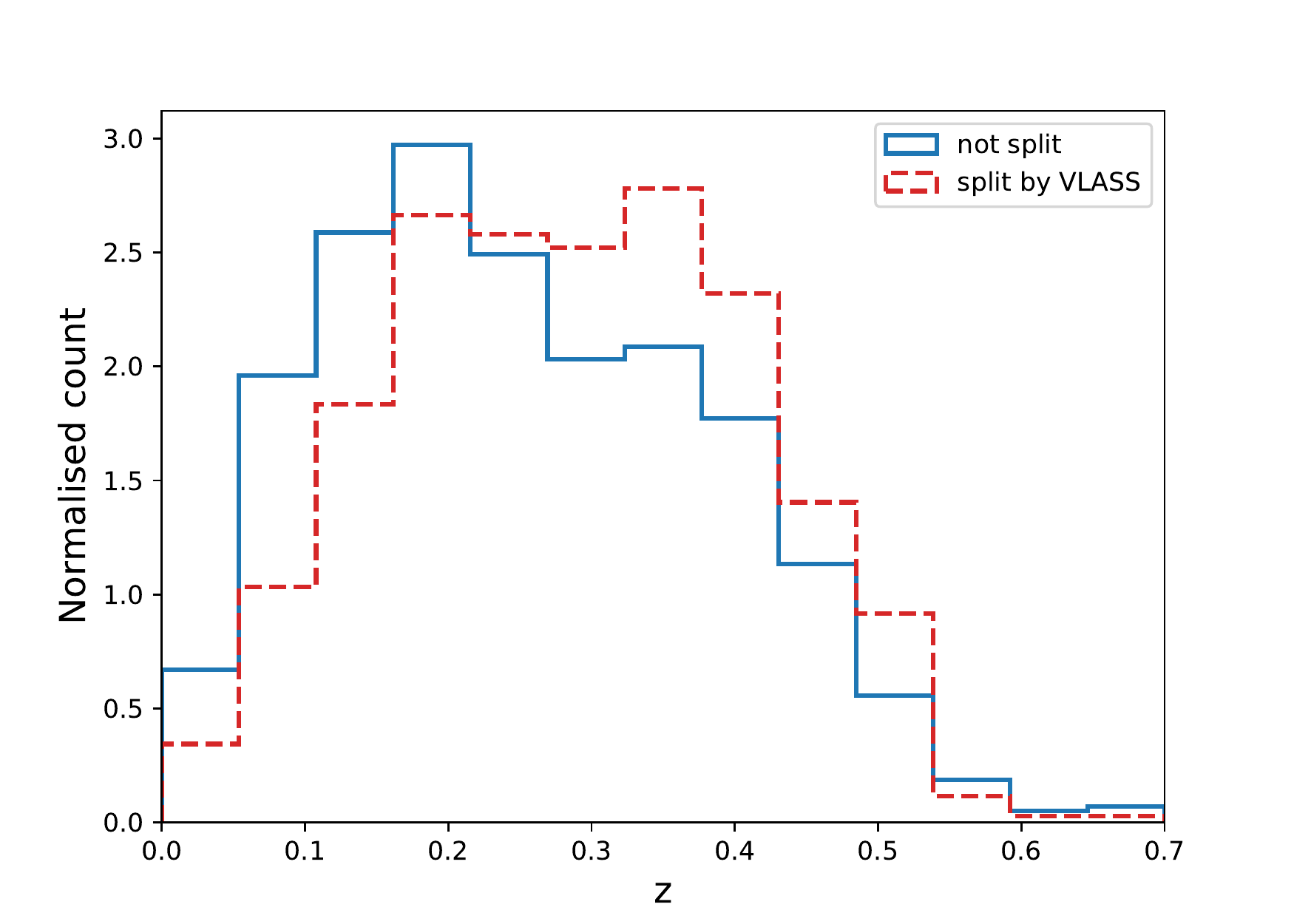}}
    \subfigure{\includegraphics[width=\columnwidth]{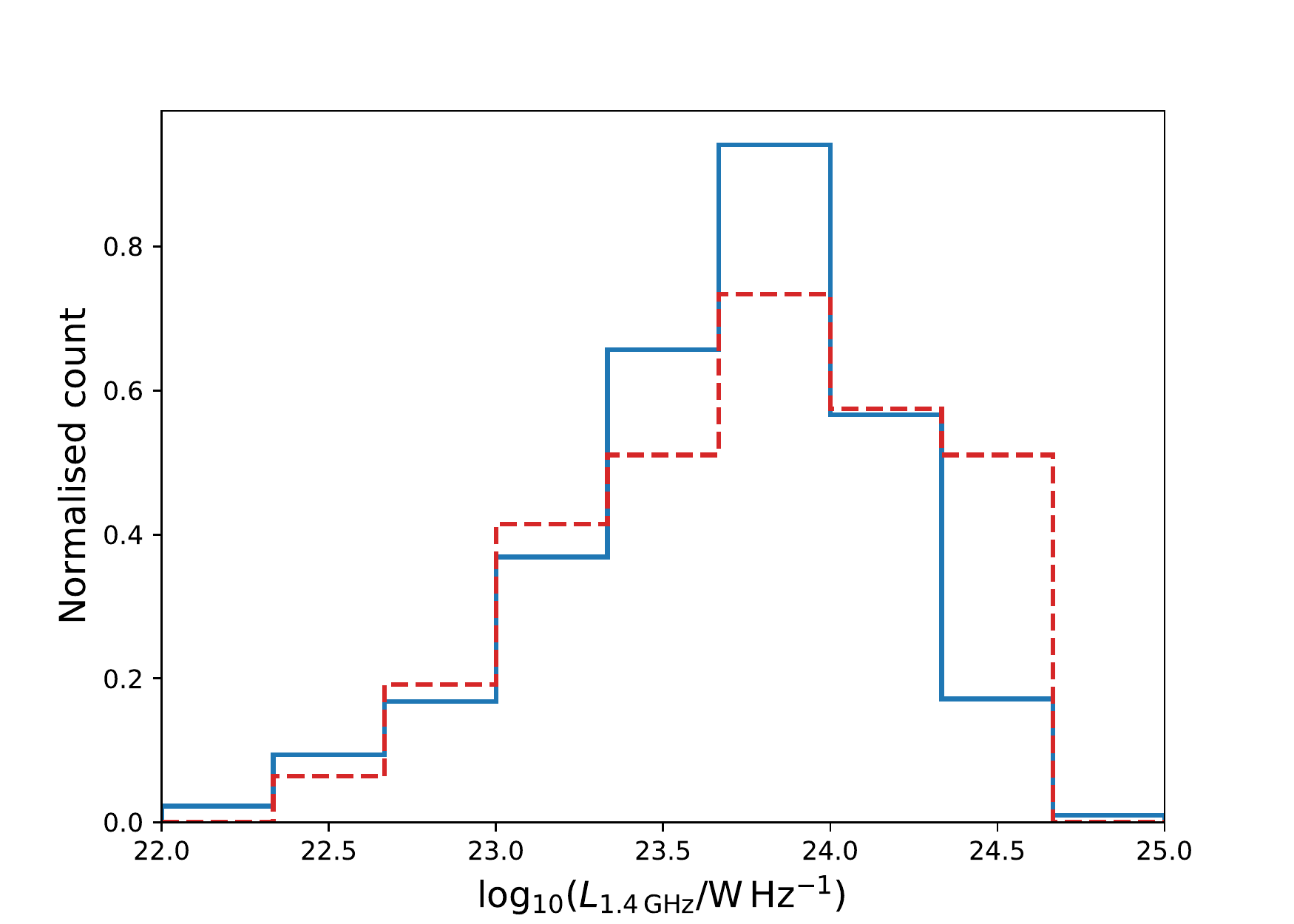}}
    \caption{The upper panel shows the normalised distributions of redshift for single-FIRST-component AGN from \citet{Best2012} with $10 < S_{1.4\,\text{GHz}} < 100\,$mJy (blue solid line), and the subset that is split into multiple components by VLASS (red dashed line).
    The lower panel shows the distributions of $1.4\,$GHz luminosity for galaxies with $z<0.15$ within these samples.} 
    \label{fig:resdists}
\end{figure}

As the \citet{Best2012} catalog includes spectroscopic data about the host galaxies, we can report on the properties of those radio galaxies split into multiple components by VLASS.
The spectroscopic properties of these sources are of interest as the AGN in Low- and High-Excitation Radio Galaxies (HERGs and LERGs respectively) are expected to powered by different accretion modes \citep[e.g.,][]{Hardcastle2006, Hardcastle2007, Buttiglione2010, Evans2011}.
However, the splitting of single-FIRST-component AGN into multiple VLASS components is the combined result of the physical extent of the radio emission and the distance to the source. 
Spectroscopic incompleteness resulting from the latter of these factors could bias the spectroscopic classification in such a comparison.
Indeed, considering only those sources with $10 < S_{1.4\,\text{GHz}} < 100\,$mJy, the redshift distribution of single-FIRST-component AGN split by VLASS differs from that of the general single-FIRST-component AGN (see Figure \ref{fig:resdists}).
Consequently, we focus on radio AGN at $z<0.15$ and $10 < S_{1.4\,\text{GHz}} < 100\,$mJy when comparing those single-FIRST-component sources split by VLASS ($n=94$, hereafter referred to as the `split' sample) to the wider single-FIRST-component AGN population ($n=927$, referred to as the `full' sample).
Moreover, taking the minimum possible separation between split components to be the VLASS beam size of $2''.5$ corresponds to a linear size of $7.5\,$kpc at a redshift of $z=0.15$.
Thus VLASS can resolve radio lobes down to galactic scales, ostensibly FR$0$s \citep{Baldi2018}, in local universe.

For the split sample $96.8_{-2.9}^{+1.0},\%$ are hosted by LERGs and $1.1_{-0.3}^{+2.4},\%$ by HERGs (with the remainder unclassified).
Comparably LERGs and HERGs account for $95.7_{-0.8}^{+0.6},\%$ and $3.2_{-0.5}^{+0.7},\%$ respectively of the full sample.
Additionally, the availability of redshift data for these objects allows us to report their radio luminosity.
In Figure \ref{fig:resdists} we show the $1.4\,$GHz luminosity distributions of the split and full sample, calculated using the flux density of the FIRST component.
The radio luminosity distributions of the two samples are statistically consistent, with a Kolmogorov-Smirnoff test returning a $p-$value of $0.06$.
The similarity in both spectroscopic classifications and radio luminosity distributions of these two samples suggests that these two populations are hosted by physically alike AGN.
This is consistent with previous studies showing radio extent alone cannot distinguish between fundamentally different populations of AGN \citep[e.g.][]{Baldi2018, Capetti2020}.


\section{Summary}
\label{sec:summary}

Using the VLASS epoch 1 \textit{Quick Look} images we have produced a catalog of radio components and characterized their quality. 
Using this we have investigated the number counts and size distributions of radio components and sources at $\nu \sim 3\,$GHz. The key points of this work are as follows:

\begin{itemize}
    \item A catalog of 1.9$\times10^{6}$ reliable and unique detections from VLASS epoch 1 \textit{Quick Look} imaging is available at \url{https://cirada.ca/catalogues}.
    Flux density measurements are underestimated by $\sim 15\,\%$ above $3\,$mJy/beam and may be less reliable below this.
    For the statistical analyses presented in this work we compensate for this by scaling our flux densities accordingly, but this correction \textit{is not} applied to the catalog data. 
    Additionally, the astrometric accuracy of this catalog is limited to $\sim 0.5''$ at Dec $>-20^{\circ}$, and $\sim 1''$ at more southerly declinations.
    
    \item Combining our catalog with FIRST shows VLASS components at $S_{3\,\text{GHz}} > 3\,$mJy/beam have a typical spectral index of $\alpha \sim -0.71$ between $1.4\,$GHz and $3\,$GHz. 
    Further combining with LoTSS DR1 demonstrates the potential for finding large numbers of radio sources with non-normal spectral curvature by utilizing upcoming survey data.
    
    \item At $S_{3\,\text{GHz}} \gtrsim 3\,$mJy the VLASS $dN/dS$ distribution is consistent with VLA-COSMOS, FIRST, and the $6$th order polynomial fit to $1.4\,$GHz observations from \citet{Hopkins2003}.
    Our VLASS catalog is shown to suffer from incompleteness at \text{$S_{3\,\text{GHz}} < 3\,$mJy.}
    
    \item The advantage of having higher angular resolution than FIRST is demonstrated via excess power in the VLASS two-point correlation function relative to FIRST at angular scales below $\sim 10''$.
    Moreover, $\sim 20\,\%$ of AGN associated with a single FIRST component are split into multiple components using VLASS.
    At $z < 0.15$, single-FIRST-component AGN that are split into multiple components by VLASS have the same radio luminosities and spectral excitation state as those that are observed as a single component by VLASS.
    
\end{itemize}

\section*{}
\authorcomment2{After the public release of our VLASS \textit{Quick Look} catalog, \citealp{Gordon2020}, and the completion of this manuscript based on that data, the authors have become aware of a second, independent, VLASS catalog that has recently been made available to the community by \citealp{Bruzewski2021}.}

\acknowledgments

The authors thank the anonymous referee for their constructive comments that have improved the quality of this work.
We thank Mark Lacy for discussions about the quick look images, and additionally JJ Kavelaars and Gilles Ferrand for their helpful feedback on the manuscript.
YAG, MMB, CPO, ANV, and SAB are supported by NSERC, the Natural Sciences and Engineering Research Council of Canada.
Partial support for LR comes from US National Science Foundation grant AST17-14205 to the University of Minnesota.
HA benefited from grant CIIC 90/2020 of Universidad de Guanajuato.

The Canadian Initiative for Radio Astronomy Data Analysis (CIRADA) is funded by a grant from the Canada Foundation for Innovation 2017 Innovation Fund (Project 35999) and by the Provinces of Ontario, British Columbia, Alberta, Manitoba and Quebec, in collaboration with the National Research Council of Canada, the US National Radio Astronomy Observatory and Australia’s Commonwealth Scientific and Industrial Research Organisation.

This research used the facilities of the Canadian Astronomy Data Centre operated by the National Research Council of Canada with the support of the Canadian Space Agency.

\vspace{10mm}

\facilities{
In this work we have made use of data obtained by the VLA, LOFAR, WISE, and the Sloan Digital Sky Survey (SDSS).
The National Radio Astronomy Observatory is a facility of the National Science Foundation operated under cooperative agreement by Associated Universities, Inc.
LOFAR data products were provided by the LOFAR Surveys Key Science project (LSKSP; \url{https://lofar-surveys.org/}) and were derived from observations with the International LOFAR Telescope (ILT). LOFAR \citep{vanHaarlem2013} is the Low Frequency Array designed and constructed by ASTRON. It has observing, data processing, and data storage facilities in several countries, which are owned by various parties (each with their own funding sources), and which are collectively operated by the ILT foundation under a joint scientific policy. The efforts of the LSKSP have benefited from funding from the European Research Council, NOVA, NWO, CNRS-INSU, the SURF Co-operative, the UK Science and Technology Funding Council and the Jülich Supercomputing Centre.
This publication makes use of data products from the Wide-field Infrared Survey Explorer, which is a joint project of the University of California, Los Angeles, and the Jet Propulsion Laboratory/California Institute of Technology, and NEOWISE \citep{Mainzer2011}, which is a project of the Jet Propulsion Laboratory/California Institute of Technology. WISE and NEOWISE are funded by the National Aeronautics and Space Administration.
Funding for the SDSS and SDSS-II has been provided by the Alfred P. Sloan Foundation, the Participating Institutions, the National Science Foundation, the U.S. Department of Energy, the National Aeronautics and Space Administration, the Japanese Monbukagakusho, the Max Planck Society, and the Higher Education Funding Council for England. The SDSS Web Site is http://www.sdss.org/.
The SDSS is managed by the Astrophysical Research Consortium for the Participating Institutions. The Participating Institutions are the American Museum of Natural History, Astrophysical Institute Potsdam, University of Basel, University of Cambridge, Case Western Reserve University, University of Chicago, Drexel University, Fermilab, the Institute for Advanced Study, the Japan Participation Group, Johns Hopkins University, the Joint Institute for Nuclear Astrophysics, the Kavli Institute for Particle Astrophysics and Cosmology, the Korean Scientist Group, the Chinese Academy of Sciences (LAMOST), Los Alamos National Laboratory, the Max-Planck-Institute for Astronomy (MPIA), the Max-Planck-Institute for Astrophysics (MPA), New Mexico State University, Ohio State University, University of Pittsburgh, University of Portsmouth, Princeton University, the United States Naval Observatory, and the University of Washington.}


\vspace{10mm}

\software{The following software packages were used in the production of this work: astropy \citep{Astropy2013, Astropy2018}, matplotlib \citep{Hunter2007}, TOPCAT \citep{Taylor2005}
}



\bibliography{VLASS_source_stats.bib}{}
\bibliographystyle{aasjournal}



\end{document}